\documentclass{aa}  
\usepackage{graphicx}
\usepackage{txfonts}
\usepackage[colorlinks=true,linkcolor=bluea,allcolors=blue]{hyperref}
\usepackage{subcaption}

\begin{document} 

   \title{The impact of relativistic AGN jets on realistic galaxy cluster environments}
    
   \author{K. Hervella-Seoane\inst{1}
          \and M. Perucho\inst{1,2}
          \and S. Planelles\inst{1,2}
          \and J.M. Mart\'{i}\inst{1,2}
          \and V. Quilis\inst{1,2}}

\titlerunning{The impact of relativistic AGN jets on realistic galaxy cluster environments}

   \institute{Departament d'Astronomia i Astrof\'{i}sica, Universitat de Val\`{e}ncia, Av/ Vicent Andr\'es Estell\'es, 19, E-46100 Burjassot, Val\`{e}ncia, Spain\\
              \email{kiara.hervella@uv.es}
         \and
             Observatori Astron\`{o}mic, Universitat de Val\`{e}ncia, C/ Catedr\`{a}tic Jos\'{e} Beltr\'{a}n 2, E-46980 Paterna, Val\`{e}ncia, Spain}

   \date{}

  \abstract
   {Low-power FRI-like radio jets dominate the active galactic nucleus population, yet the main mechanism of heating in the intracluster medium (ICM) and its efficiency remain a matter of debate.}
   {We investigate the impact of intermediate-power FRI-like relativistic jets on the inner region of a realistic 
   ICM, focusing on the role of weak shocks in heating and the environmental coupling in regulating energy deposition.}
   {We present six 3D relativistic-hydrodynamic simulations of ($10^{44}-10^{45}$) erg/s FRI-like jets propagating through the central regions of realistic ICM environments extracted from cosmological \textsc{GADGET-3} runs. Each simulation tracks 20 Myr of continuous jet activity with radiative cooling included.}
   {Despite their intermediate power and mildly relativistic speed at injection, the jets rapidly decelerate to transonic velocities (average Mach numbers of $\sim 2.3$) and generate weak bow shocks that dominate energy transfer. Approximately 80 $\%$ of the injected kinetic power is converted into the internal energy of the ambient ICM, and bremsstrahlung cooling noticeably reduces the temperature of the shocked shells as they expand.}
    {Our results demonstrate that the heating of the ICM and intergalactic medium is highly efficient even in the case of weak shocks in FRI-like sources. This finding suggests that low-power jets are a viable solution to the long-standing cooling-flow problem and likely regulate star formation across diverse galaxy cluster environments.}

    \maketitle
   
\section{Introduction}\label{introduction}

Extragalactic jets are usually observed in massive elliptical galaxies, both isolated galaxies and those in rich clusters. These streams of plasma are generated in the surroundings of the central supermassive black holes as a result of accretion, and they likely form when the magnetic field extracts rotational energy from the supermassive black hole, as described by \citet{blandford_electromagnetic_1977} analytically and tested in multiple numerical simulations \citep{tchekhovskoy_efficient_2011,tchekhovskoy_prograde_2012,porth_three-dimensional_2013,porth_black_2017,talbot_blandfordznajek_2022,fromm_impact_2022}. 
Over the last decade, radio observations have allowed us to resolve the innermost region of some jets \citep{hada_origin_2011,hada_high-sensitivity_2016,baczko_highly_2016,collaboration_first_2019,boccardi_jet_2021,ricci_spectral_2025}. These observations have shown that jets collimate within the first $10^3 - 10^6$ gravitational radii. Then, jets are accelerated up to relativistic speeds by thermal and magnetic mechanisms \citep{komissarov_magnetic_2007,lister_mojave_2009,homan_mojave_2015,ricci_magnetic_2024} within the inner few kiloparsecs, and they propagate up to $\sim$ 30 kpc, interacting strongly with their galactic and cluster environments.

At kiloparsec scales, extragalactic jets are mainly observed at radio frequencies, and they show a morphological and brightness dichotomy \citep{fanaroff_morphology_1974} in between those of powerful Fanaroff-Riley class II (FRII) sources, which have bright hotspots and lobes, and weaker FRI sources, which have irregular and de-collimated large-scale structures. Historically, jet power has been considered to be the main factor responsible for this dichotomy \citep{rawlings_evidence_1991,ghisellini_dividing_2001}, with a transition regime around $10^{44}$ erg/s. Multiple studies have also shown that jet morphology can be a result of the complex interplay between the jet's initial energetics and the surrounding medium \citep{gopal-krishna_fanaroff-riley_2001,croston_environments_2019,mingo_revisiting_2019}, in particular at intermediate powers, i.e., in the power transition between FRI and FRII. In the case of FRI sources, de-collimation is probably caused by mass entrainment and the turbulent mixing of the flows within the host interstellar medium \citep[ISM; e.g.,][]{bicknell_relativistic_1995}, including stellar winds \citep[][]{komissarov_mass-loaded_1994,perucho_large-scale_2014,angles-castillo_deceleration_2021}. Mixing with the ISM is probably driven by small-scale instabilities arising at the jet boundary, as suggested by observational work \citep{laing_systematic_2014}. The different possible factors responsible for this include the Rayleigh-Taylor instability, the centrifugal instability \citep{matsumoto_linear_2017,gourgouliatos_reconfinement_2018}, and Kelvin-Helmholtz short wavelength modes \citep{perucho_stability_2010}. It has been proposed that stars piercing and penetrating the jet along their orbital trajectories results in an initial perturbation that excites the growth of unstable modes \citep{perucho_triggering_2020}. Although stability theory shows that low-power jets are significantly more susceptible to disruption and long-wavelength instabilities develop in numerical simulations \citep[e.g.,][]{massaglia_making_2016,tchekhovskoy_three-dimensional_2016}, radio maps of FRI radio galaxies do not show these features, indicating that a small-scale process is responsible for mass-load and deceleration between 1 and 3 kpc from the central engine \citep{bicknell_model_1984,laing_systematic_2014}. These low-power jets are finally decelerated to sub-relativistic, transonic velocities, thus damping Doppler boosting; they show notable symmetry at kiloparsec scales. Consequently, the transition region in power represents a regime where environmental properties play a decisive role in determining whether a jet can sustain its collimation. In contrast, FRII jets remain fairly collimated and mildly relativistic at large scales, as indicated by the brightness asymmetry between the jet and counter-jet. 
 
Observational support for active galactic nucleus (AGN) feedback has grown significantly over the past two decades: X-ray \textit{Chandra} and \textit{XMM–Newton} observations have revealed a lack of cool gas in the centers of many galaxy clusters, contrary to expectations based on radiative cooling times. This discrepancy, known as the cooling-flow problem, strongly suggests that an additional heating mechanism that can offset cooling must be operating. Galactic nuclear activity provides the most plausible solution, and evidence of its impact has been confirmed across a wide range of scales and environments, through multiple physical processes (see \citealt{harrison_observational_2024}, \citealt{mukherjee_jet_2025}, and references therein). Radio jets are one of the main manifestations of nuclear activity in galaxies and are widely recognized as a key process in galaxy formation and evolution. By heating, displacing, and removing the gas that would otherwise fuel star formation, AGN feedback can potentially regulate black hole and galaxy growth \citep{marconi_local_2004}, preventing galaxies from becoming over-massive \citep{di_matteo_energy_2005,vogelsberger_introducing_2014,somerville_physical_2015,dubois_horizon-agn_2016} and thereby reproducing the observed galaxy--halo mass relations \citep{silk_quasars_1998,croton_many_2006}. Population studies of radio-loud AGNs  \cite[e.g.,][]{best_fundamental_2012} that consider the radio luminosity as a proxy for jet power \citep[following, e.g.,][]{cavagnolo_relationship_2010} have shown that low-power AGNs make up the majority of the AGN population \citep{hardcastle_radio-loud_2019}. As such, they are key for understanding the impact and relevance of AGN feedback in galaxy formation and evolution \citep{heckman_coevolution_2014,sabater_lotss_2019}. 

One of the most compelling pieces of observational evidence of AGN jet feedback is the X-ray cavities found in galaxy groups and clusters, which coincide with the jet-inflated radio lobes \citep{fabian_very_2006,mcnamara_heating_2007}. The displacement of hot gas by these cavities implies mechanical work done by the jet on the intracluster medium (ICM); this allows researchers to derive AGN jet powers, and they are consistent with those estimated from radio observations. However, this buoyancy-driven cavity model is no longer sufficient to fully explain the observations, since radio lobes have been found to be surrounded by weak shocks (Mach numbers of $\approx 1 - 2$) in many systems \citep{mcnamara_heating_2005,simionescu_large-scale_2009,croston_large-scale_2011,stawarz_interaction_2014}. This indicates that the jet lobe expansion must still be pressure-driven rather than rising buoyantly and that shocks play a critical role in lobe inflation. Simulations require a relativistic approach for an appropriate description of this shock-driven phase, which allows a significant fraction of the jet's internal and kinetic energy to be exchanged with the ambient \citep{perucho_radio_2017}. This fact, along with the shorter timescales implied by strong shocks mediating the energy exchange as shown by multiple simulations \citep[e.g.,][]{perucho_intracluster_2011,perucho_large-scale_2014, perucho_long-term_2019, wagner_relativistic_2011,matsumoto_propagation_2019, rossi_different_2020}, makes them extremely efficient at heating the galactic and cluster gas. 

However, it is also worth mentioning that typical observations of jet feedback in galaxy clusters \citep{fabian_very_2006,croston_large-scale_2011,sanders_very_2016} and simulations \citep{bruggen_shock_2007,li_modeling_2014,bourne_agn_2021,ehlert_self-regulated_2022} have found that shocks are often not very strong. Strong shocks might be more relevant in the early stages of evolution of all sources and throughout the whole evolution in the case of FRII jets; weak shocks, on the other hand, are more common on larger scales, outside the immediate jet and lobe region, and in the case of FRI or low-power jets. While strong shocks from powerful FRII jets have been extensively studied \citep{perucho_interaction_2010,perucho_large-scale_2014, perucho_long-term_2019, wagner_relativistic_2011,matsumoto_propagation_2019, rossi_different_2020}, the weak shocks characteristic of FRI-like sources, which dominate the radio-loud AGN population, remain poorly understood in terms of their heating impact.

On top of that, the main dominant heating mechanism (raising bubbles, shocks, turbulence, etc.) can vary depending on jet power and coupling with the ambient gas. Numerical simulations \citep{mukherjee_relativistic_2018,perucho_long-term_2019,meenakshi_modelling_2022,bourne_recent_2023} have shown that these factors strongly influence how efficiently jet energy is deposited into the ISM  and ICM. Furthermore, the environment (e.g., the density profile, cluster core structure, etc.) is relevant to jet evolution. Jets simulated in an idealized and smooth ICM behave differently than when the cluster presents substructures. We refer the reader to \citet{bourne_recent_2023} and \citet{mukherjee_jet_2025} for up-to-date reviews on the effort made through simulations to study jet feedback mechanisms at different scales and the different environments introduced in simulations.

In this work, we present six different 3D numerical simulations of relativistic jets in four different environments. We injected multiple jets (with kinetic powers of $10^{44}-10^{45}\,{\rm erg/s}$) into different galaxy clusters obtained from \textsc{GADGET}-3 cosmological runs. Even with the intermediate power of such jets, their mildly relativistic velocities at injection reproduced the morphologies of FRI sources. In this way, we studied the heating mechanisms in jets within this parameter range in realistic cluster environments, obtained from cosmological simulations, and their effect on jet propagation and morphology.

In Sect. \ref{sec_simulations} we describe the computational setup for the simulations. Section \ref{sec_results} includes the results, focusing on the jets' propagation and evolution as well as the developed large-scale structures and their impact on the heating of the ICM. In Sect. \ref{sec_discussion} we discuss these results and compare them with previous works. Finally, in Sect. \ref{sec_conclusions} we summarize the findings of our work.

\section{Simulations}\label{sec_simulations}

\subsection{Computational setup}\label{sec_cosmology}

For the simulations presented in this paper, we used the latest version of the code \textsc{Ratpenat} \citep{perucho_interaction_2010,perucho_long-term_2019}, a hybrid parallel code (\textsc{mpi}+\textsc{openmp}) that solves the equations of relativistic hydrodynamics (RHD) in conservation form through high-resolution-shock-capturing methods: (i) primitive variables within numerical cells are reconstructed using piecewise parabolic method routines, (ii) numerical fluxes across cell interfaces are computed with the Marquina flux formula, and (iii) the advance in time is performed with third-order total variation diminishing preserving Runge–Kutta methods. 

The equations that are solved, in Cartesian coordinates, are those corresponding to the conservation of mass, momentum and energy \cite[a detailed description of the equations is presented in][]{perucho_long-term_2019}. Cooling is taken into account following the approximation given by \citet{myasnikov_radiative_1998} and included as a source term in the energy equation, implicitly assuming thermodynamical equilibrium in each cell. We also assumed that thermal cooling only affects the post-shock ambient medium because it is only relevant in dense, nonrelativistic (NR) regions, which implies that losses can be considered isotropically radiated (the medium is optically thin), and thus cooling only needs to be considered in the energy equation \citep{perucho_radio_2017}. Finally, the system is closed by the Synge equation of state \citep{synge_relativistic_1957}, which is solved for a mixture of relativistic Boltzmann gases, including populations of electrons, positrons, and protons (see \citealt{perucho_numerical_2007} for more details regarding the implementation).
The code also integrates an equation for the jet mass fraction that is used for tracing the jet material through the grid. 

For the main set of simulations the 3D numerical grid contains 512 cells in the $y$ axis, the jet axis direction, and 768 in each transversal direction; with a resolution of 125 pc/cell. This translates to a total physical grid size of $(96 \times 64 \times 96)$ kpc. For the grid boundary conditions, we imposed reflection at the jet base in order to mimic the presence of a counter-jet, and outflow in the rest of directions.

All these simulations were performed in Mare Nostrum, at the Barcelona Supercomputing Center, part of the Red Española de Supercomputación (Spanish Supercomputing Network), under the projects AECT-2024-3-0011: ``The complex dance between relativistic outflows and their host galaxies.'' These simulations required a total of $\sim 4$ million hours of computing time.

\subsection{Ambient medium: Galaxy clusters}\label{sec_clusters}

\begin{table}
\caption{Main properties of the four galaxy clusters, at $z = 0$. }
\centering
\begin{tabular}{cccc} \hline
{Cluster} & {$M_{\rm vir}$} & {$R_{\rm vir}$} & {$R_{200}$}\\ 
& ($10^{14}\,h^{-1}\,M_{\odot}$) & ($h^{-1}$ Mpc) & ($h^{-1}$ Mpc)\\ \hline
{D02} & 5.26 & 1.70 & 1.19\\
{D03} & 6.53 & 1.83 & 1.28\\
{D06} & 15.41 & 2.43 & 1.69\\
{D10} & 15.46 & 2.43 & 1.65\\ \hline
\end{tabular}
\label{tab_clusters}
\tablefoot{From left to right, the columns report the cluster ID, virial mass, virial radius, and $R_{200}$ radius.}
\end{table} 

\begin{figure}[htb]
    \includegraphics[width=0.45\textwidth]{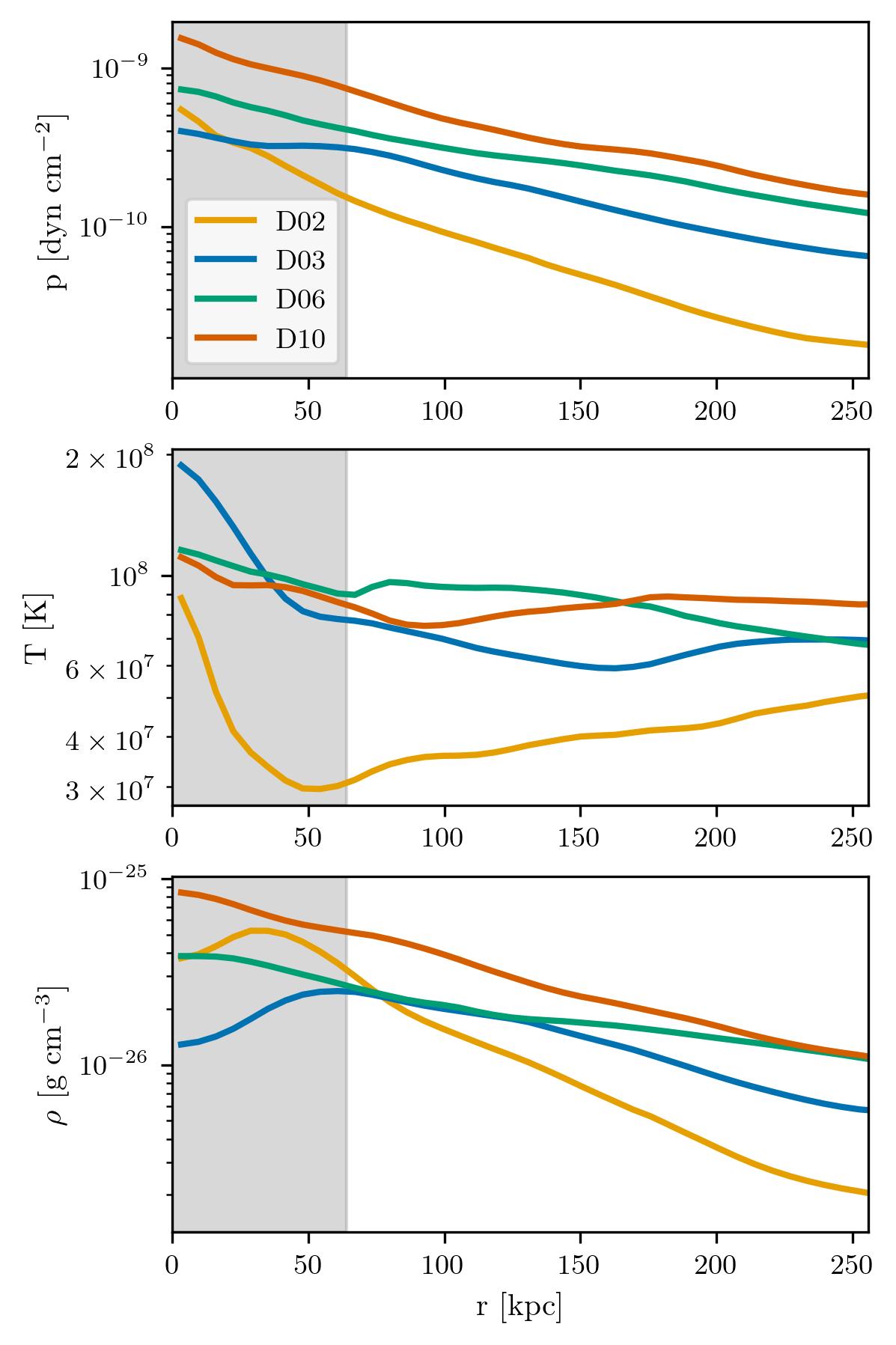}
\caption{Pressure, density, and temperature average radial profiles of the galaxy cluster environments used in our simulations. The gray-shaded region denotes the central 64 kpc domain where jets are injected. These profiles are the result of a direct spatial interpolation from the original clusters presented in \cite{planelles_role_2014}. }
\label{Dianoga_profiles}
\end{figure}

The jets were injected into a computational grid filled with the inner region of different galaxy clusters from the cosmological hydrodynamic simulations presented in \citet{planelles_role_2014}. The original sample consists of zoomed-in simulations of massive dark-matter haloes extracted from the lower-resolution $N$-body parent cosmological simulation in \cite{bonafede_non-ideal_2011}. The cosmology assumed corresponds to a flat $\Lambda$ cold dark matter cosmological model, with matter density parameter $\Omega_{\rm m}$ = 0.24, baryon density parameter $\Omega_{\rm b}$ = 0.04 and Hubble constant 
$H_0 = 72\,h$ km s$^{-1}$ Mpc$^{-1}$.
As explained in \cite{planelles_role_2014}, each region has been re-simulated with the \textsc{TREEPM–SPH GADGET}-3 code, originally from \citet{springel_cosmological_2005}, increasing the mass resolution and including the baryonic component. For the hydrodynamical description, the authors use a smoothed particle hydrodynamic model \citep{beck_improved_2015}, with a resolution for the dark matter particles of $m_{\rm DM} = 8.47 \times10^8 \,h^{-1} \, M_{\odot}$ and of $m_{\rm gas} = 1.53 \times 10 ^8 \, h^{-1} \, M_{\odot}$ for the gas particles. These re-simulations include a large variety of physical processes to describe the evolution of the gas (the included baryonic component): heating/cooling from both the cosmic microwave and the UV time-dependent backgrounds \citep{haardt_modelling_2001}, metallicity-dependent radiative cooling \citep{wiersma_effect_2009}, star formation \citep{springel_cosmological_2003} and the corresponding metal enrichment \citep{tornatore_simulating_2004}, supernova feedback \citep{springel_cosmological_2003} and a black hole accretion with an AGN feedback model \citep{steinborn_refined_2015}. For further details and other thermo- and chemodynamical properties of these re-simulations, we refer the reader to a series of papers \citep{planelles_role_2014,rasia_cool_2015,biffi_history_2017,planelles_pressure_2017,truong_cosmological_2018,biffi_origin_2018} .

The systems included in the present work constitute a representative subsample of four galaxy clusters, chosen to include two massive (D6 and D10) and two smaller systems (D2 and D3); each system comprises one cool core and one non-cool core. Their cool-coreness has been studied by \citet{rasia_cool_2015} and classified based on their pseudo-entropy inner-region value and profile. The global properties of the four galaxy clusters, namely the virial masses and radii, and the $R_{200}$ radii  are listed in Table~\ref{tab_clusters}.

For our simulations, we considered the environment to be static, using just one $z$-fixed snapshot for each environment (at $z = 0$). This assumption comes from the difference between the timescale for jet evolution considered in this paper, of around 20 Myr, as compared to the characteristic timescales for noticeable ICM changes in their structure. Given that typical velocities for the ICM gas are around at $v_{\rm{ICM}} = 10^3 $~km/s \citep{simionescu_constraining_2019} and, for our clusters, $R_{\rm{vir}} \sim 1 $~Mpc, we obtain a dynamical timescale of $\Delta t_{\rm{ICM}} \sim $ 1 Gyr. Hence, we do not expect significant changes in the gas distribution during the considered jet injection and evolution time. In order to keep the ambient medium in equilibrium, a restoring force (an external gravity) is included as source terms in the momentum and energy hydrodynamical equations, as it was done in previous works \citep[e.g.,][]{gomez_hydrodynamical_1997,perucho_numerical_2007,perucho_interaction_2010,perucho_large-scale_2014,perucho_long-term_2019}. 

For our jet simulations, we restricted the computational domain to the central $(96 \times 64 \times 96)$ kpc, corresponding to the inner ICM (also referred to as the central regions of the galaxy cluster), formed by the central cluster galaxy and its surroundings. Throughout this work, references to the ICM or ambient medium pertain exclusively to this core region. In order to adapt the cosmological snapshots as environments for the RHD code \textsc{Ratpenat}, a few adjustments were necessary. First of all, given that the cosmological spatial scales vastly exceed the estimated jet radii from observations, we performed a direct interpolation of the chosen $z$-fixed snapshot for each environment to achieve the 125 pc/cell. This corresponds to a resolution at injection of 4 cells per jet radius, which is sufficient to accurately capture the overall jet structure and its interaction with the surrounding medium at large scales, i.e., the main focus of this study. Second, for consistency with \textsc{Ratpenat}, the ambient medium was required to satisfy the Synge equation of state. To achieve this, we used the pressure and temperature values from the simulations as input and allow the code to compute the corresponding densities. These densities differ slightly from the original ones but still preserve the same substructures. The resulting pressure, density and temperature average radial profiles of each galaxy cluster environment are shown in Fig. \ref{Dianoga_profiles}.

\subsection{Jet parameters}\label{sec_jets}

\begin{table*}
\caption{Parameters of the simulated jets.}
\centering
\begin{tabular}{cccccccccc} \hline
Simulation ID & Environment &  Resolution & $L_{\rm j}$ & $r_{\rm j}$ & $v_{\rm j}$ & $\rho_{\rm j}$& $\rho_{\rm j}/\rho_{\rm a,0}$ & $x_e$ & $T_{\rm j}$ \\
 &  & [pc/cell] & [erg $\rm{s^{-1}}$] & [kpc] & [c]  & [g $\rm{cm^{-3}}$] & & & [K] \\ \hline
J3C3\_144 & D03 & 125 & $10^{44}$ & 0.5 & 0.4 & $2.7\times10^{-31}$ & $1.4\times10^{-5}$ & 0.5 & $2\times10^{10}$\\ 
J3C3\_544 & D03 & 125 & $5\times10^{44}$ & 0.5 & 0.4 & $7.5\times10^{-31}$ & $4.0\times10^{-5}$ & 0.5 & $2\times10^{10}$\\ 
J3C2\_544 & D02 & 125 & $5\times10^{44}$ & 0.5 & 0.4 & $7.5\times10^{-31}$ & $8.5\times10^{-6}$ & 0.5 & $2\times10^{10}$\\
J3C6\_544 & D06 & 125 & $5\times10^{44}$ & 0.5 & 0.4 & $7.5\times10^{-31}$ & $1.4\times10^{-5}$ & 0.5 & $2\times10^{10}$\\
J3C6\_145 & D06 & 125 & $10^{45}$ & 0.5 & 0.6 & $1.0\times10^{-30}$ & $1.9\times10^{-5}$ & 0.5 & $2\times10^{10}$\\  
J3C1\_145 & D10 & 125 & $10^{45}$ & 0.5 & 0.6 & $1.0\times10^{-30}$ & $1.2\times10^{-5}$ & 0.5 & $2\times10^{10}$\\
\hline
\end{tabular}
\label{tab_simulations}
\tablefoot{Columns list (from left to right): the simulation ID, the galaxy cluster environment, numerical resolution, jet power, injection radius, injection velocity, injection density and jet-to-ambient density ratio at injection, leptonic number fraction, and jet temperature.}
\end{table*} 

We performed a total of six 3D RHD simulations of mildly relativistic jets propagating through the four different environments described in the previous section.
The injected jets have kinetic powers in the range of $L_{\rm j} = 10^{44} - 10^{45}\,{\rm erg/s}$, as shown in Table~\ref{tab_simulations} together with other jet parameters used in each simulation. Jets with lower power could not be reliably simulated due to combined numerical and physical constraints. First, although interpolation was applied to the cosmological environments, further increasing the resolution would artificially distort the environment properties, which sets a minimum jet radius for our models. Second, we imposed the physical requirement that the jets be at least in pressure equilibrium with the ambient medium to ensure stable propagation. Finally, we constrained the jet velocity and density to values consistent with observations and previous simulation results at these injection distances, where the jet must be already slightly mass-loaded and decelerated \citep{angles-castillo_deceleration_2021}. All these constraints, either guided by simulation limitations or motivated by realistic physical conditions, limit the possible jet powers for the chosen environments.

The jets are injected for a duration of $\approx$ 20 Myr at 2kpc from the alleged center of its host galaxy cluster. They have a radius $R\rm{_j} = 0.5$~kpc at injection, implying a half-opening angle of 14$^\circ$, consistent with observations of well-known FRI sources \citep{laing_systematic_2014}. The injection jet temperature corresponds to $2\times 10^{10}$~K, which implies that the jets are thermodynamically relativistic (with a specific enthalpy $h \approx 8\,c^2$). 

In all cases, the jet densities at injection range between $\rm{\rho_j} \approx 10^{-32}$ g cm$^{-3}$ and $\rm{\rho_j} \approx 10^{-30}$ g cm$^{-3}$. These values are extrapolated from the simulations reported by \cite{angles-castillo_deceleration_2021}, where mass-load by stellar winds within the two inner kiloparsecs is taken into consideration. This results in a density ratio between the jet material and the environment of $\rm{\rho_j/\rho_{a,0}} \approx 10^{-5}-10^{-6}$. 

Given our jet injection position, and the mass-load, we assumed the FRI-like jets to have already decelerated into mildly relativistic speeds. Hence, jets are injected along the jet propagation axis, with velocities ${v_y} = {v_{\rm j}} =  (0.4 - 0.6)$ c, and transverse velocities in the $x$-$z$ plane described as a constant-magnitude radial perturbation,
\begin{equation}
    v_x = \frac{0.1 \, v_{\rm j}}{\sqrt{2}} \sin \theta, \qquad v_z = \frac{0.1 \, v_{\rm j}}{\sqrt{2}} \cos \theta,
\end{equation}

\noindent where $\theta = \arctan(z/x)$. These terms introduce the effect of jet expansion at the base, to make it consistent with the jet radius at injection. 

\section{Results}\label{sec_results}

\subsection{Evolution}\label{sec_results_evolution}

\begin{figure}[htbp]
    \includegraphics[width=0.47\textwidth]{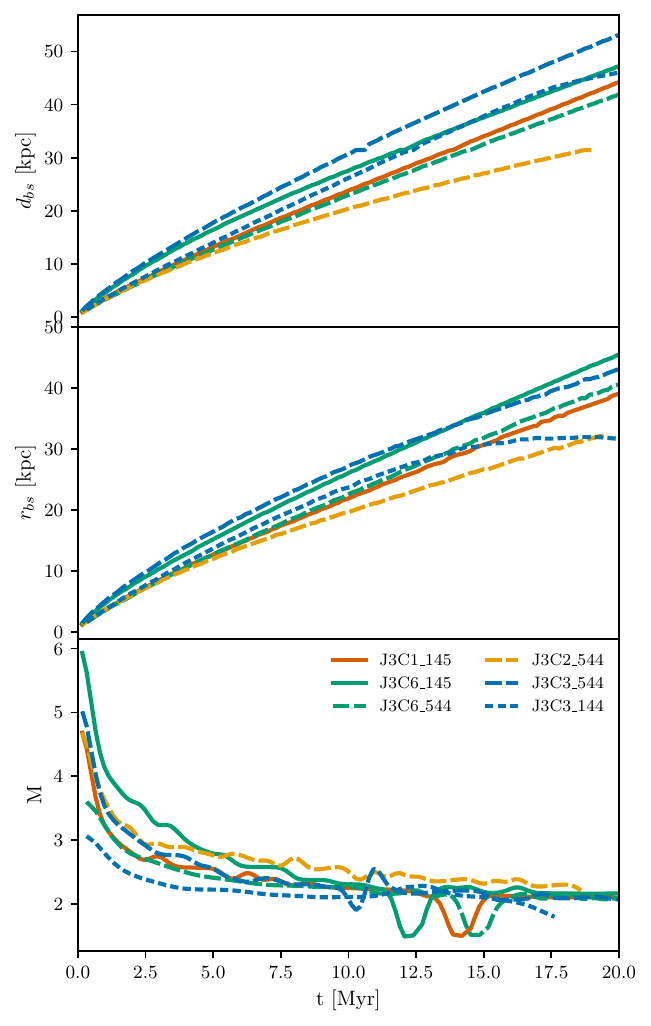}
\caption{Time evolution of the bow shock tip position (\textit{top}), bow shock maximum radius (\textit{center}),  and Mach number of the bow shock (\textit{bottom}). The simulation names are presented in Table \ref{tab_simulations}. The colors indicate the different environments: orange for D10, green for D06, blue for D03, and yellow for D02. The line styles indicate different jet powers simulated in those environments: solid lines for jets with a power of $10^{45}$ erg/s, dashed for $5\times10^{44}$ erg/s, and narrow dashes for $10^{44}$ erg/s. The Mach number lines have been smoothed for clarity.}
\label{shock}
\end{figure}

\begin{figure}[htbp]
    \includegraphics[width=0.47\textwidth]{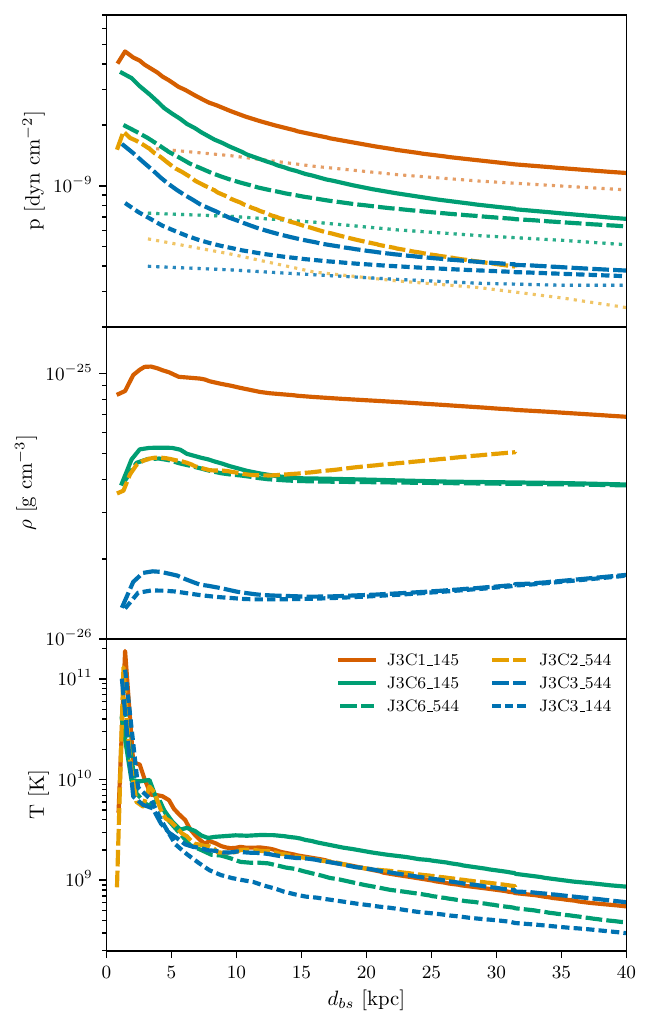}
\caption{Pressures (\textit{top}), densities (\textit{center}), and temperatures (\textit{bottom}) of the shocked ICM (in cgs units) versus the position of the bow shock (in kpc). Colors are used to indicate the different environments of each simulation, and the line styles indicate the jet power, as in Fig. \ref{shock}. The dotted lines in the pressure panel correspond to the environment radial profiles.}
\label{shocked_ICM}
\end{figure}

\begin{figure}[htbp]
    \includegraphics[width=0.47\textwidth]{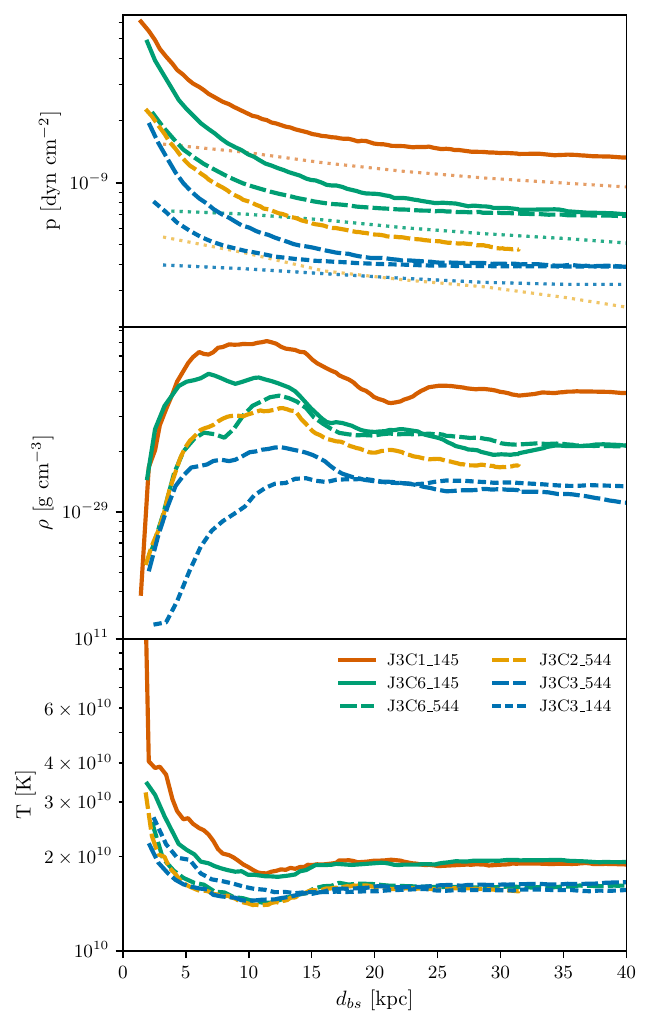}
\caption{Cocoon pressures (\textit{top}), densities (\textit{center}), and temperatures (\textit{bottom}), in cgs units, versus the position of the bow shock, in kpc. Colors and line styles are the same as in Figs. \ref{shock} and \ref{shocked_ICM}.}
\label{cocoon}
\end{figure} 

All our simulations follow the active phase of the jets during 20 Myr of continuous injection. All the jets follow the same general characteristic evolution, developing similar morphological features (to be analyzed in sect.~\ref{sec_results_structure}): a weak bow shock produced by the supersonic flow propagating into the ambient medium, a jet terminal shock where the flow is strongly decelerated, and a cocoon inflated by the shocked jet flow and polluted by the ICM shocked gas (via mixing at the contact discontinuity between the two media). The region between the bow shock and the cocoon (the shell of the cavity carved by the jet) is formed mainly by shocked ambient medium and has a large density.

The analysis of the jet evolution begins with the identification of the bow shock during post-processing, using thresholds based on pressure jumps and velocity. In practice, the shock is detected by: (a) the first increase (measured from the outer boundary of the computational domain inward) in the local ambient pressure exceeding $5\%$; and (b) assuming that the ambient medium is in hydrostatic equilibrium, the presence of a velocity significantly larger than any possible numerical noise (we adopted a minimum threshold of $10^{-4}\,c$). Together, these two criteria allowed us to unambiguously identify weak shocks in our simulations, avoiding confusion with static substructures in the ambient medium.

The top panel of Fig.~\ref{shock} displays the position of the shock head, defined as the farthest position of the bow shock along the main propagation direction. When comparing jets with same power and properties in different environments (e.g., J3C3$\_$544, J3C6$\_$544, and J3C2$\_$544, represented by dashed lines), we find that the axial distances reached by their bow shocks anticorrelate with the clusters average densities (see the D03, D06, and D02 cluster profiles in Fig. \ref{Dianoga_profiles}). The central panel shows the maximum radial position of the bow-shock. Given the similarity between the axial and radial distances, we can state that the bow shocks developed by the jets expand quasi-spherically throughout the whole simulation. In the bottom panel, we show the estimated Mach numbers ($M$) for the axial propagation of the bow shocks (computed as the instantaneous bow shock propagation velocity  relative to the ambient sound speed), with initial values between 3 and 6, but average values falling to $\approx 2.3$, indicating a transition toward the transonic regime.

\begin{figure*}[htbp]
    \includegraphics[width=\textwidth]{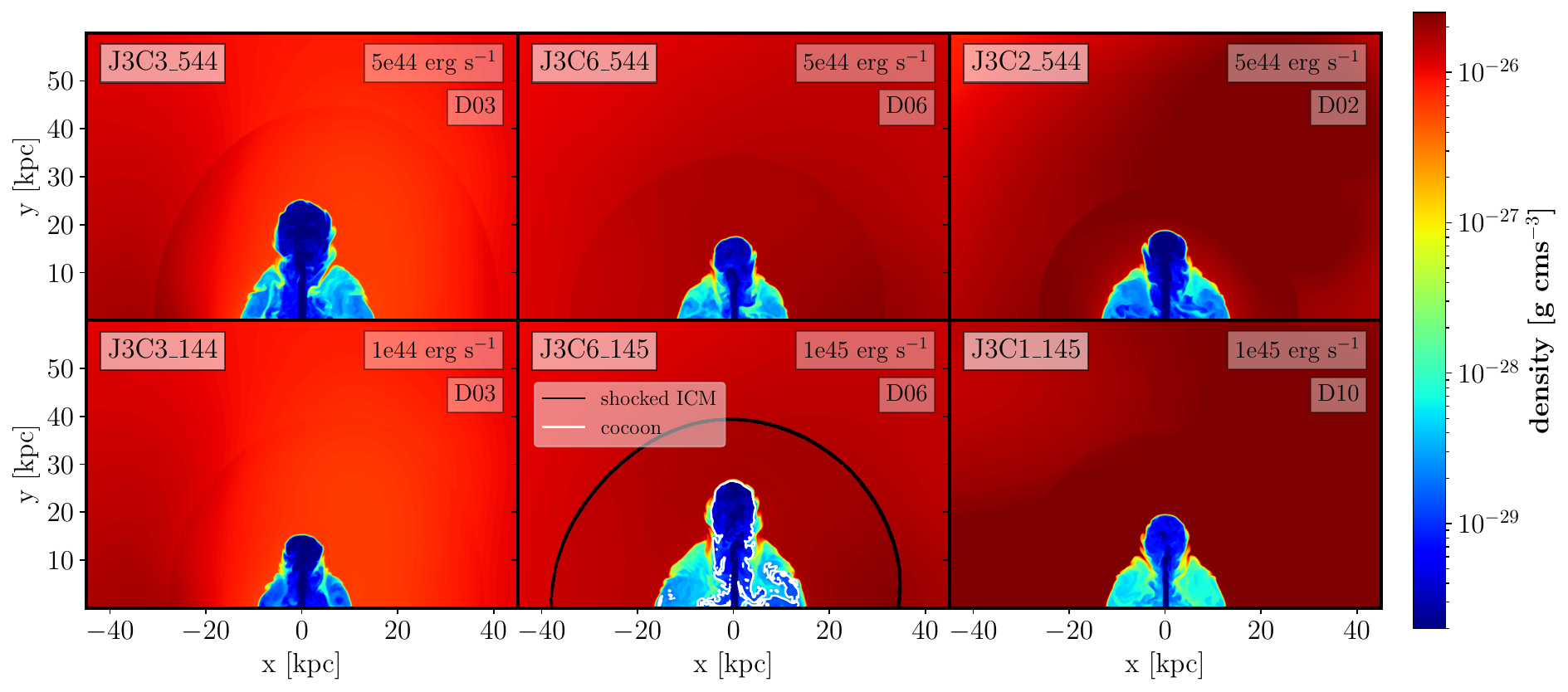}
\caption{Density maps of each simulation at the dynamical time t = 16 Myr. \textit{Top row}: Simulations of $5\times 10^{44}\,{\rm erg/s}$ jets in the environments D03 (left), D06 (center), and D02 (right). \textit{Bottom row}: Other jet power simulations for D03 (left) and D06 (center) environments and the $10^{45}\,{\rm erg/s}$ simulation in D10 (right). The black and white contours shown in the central-bottom panel trace the shocked ICM shell and cocoon, respectively, for the simulation J3C6$\_$145.}
\label{Fig_dens}
\end{figure*}

We then identified the different jet components using a particle tracer that represents the relative mass fraction of jet material in a given cell. Inside the shocked region, we identified the cocoon (including the jet) with the domain with tracer values between $10^{-2}$ and 1, and the shocked ICM (the shell) with that with tracer values below $10^{-2}$ 
(see the contours for the J3C6$\_$145 simulation in the bottom-center panel of Fig. \ref{Fig_dens}). For both cocoon and shell regions, we computed key physical magnitudes throughout the entire simulation. Figures~\ref{shocked_ICM} and \ref{cocoon} show the evolution of the volume-averaged pressure (top panels), density (central panels), and temperature (bottom panels) for the shocked ICM and the cocoon, respectively. We can observe how, for all simulations, both the shell and the cocoon start significantly over-pressured relative to environment, but, as time evolves, the shell pressure approaches that of the unshocked medium. The cocoon density increases at the beginning of the injection, as the under-dense injected material mixes with the ambient gas. However, as it evolves, the expansion leads to a decrease. The shell density evolution shows different behaviors, decreasing as the shock expands for most simulations. However, for those jets evolving through the D02 and D03 clusters, an initial drop in density is followed by an increase, showing a strong coupling between the shell and the environmental densities (see the density profiles in Fig.~\ref{Dianoga_profiles}). 
Regarding temperatures, we observe an early fast decline in both the cocoon and the shell during the initial stages of the evolution. Subsequently, the cocoon temperature remains approximately constant throughout the simulation. In contrast, the shocked ICM exhibits a continuous temperature decline, driven by radiative cooling losses. Cooling does not significantly affect the cocoon owing to its low densities. The impact of cooling on the evolution of the shocked ambient medium was not apparent in previous studies of more powerful jets, where stronger shocks dominated the thermal evolution \citep{perucho_long-term_2021}.

Finally, we must mention that we do not find any connection between the jet large-scale morphology and average propagation distances and the cool-coreness of their host galaxy clusters. This is expected, given that this property involves larger spatial scales than those reached by our simulated jets. In the inner tens of kiloparsecs, the host galaxies and their immediate intergalactic medium profiles are similar, regardless of the temperature profile.

\begin{figure*}[htbp]
    \includegraphics[width=\textwidth]{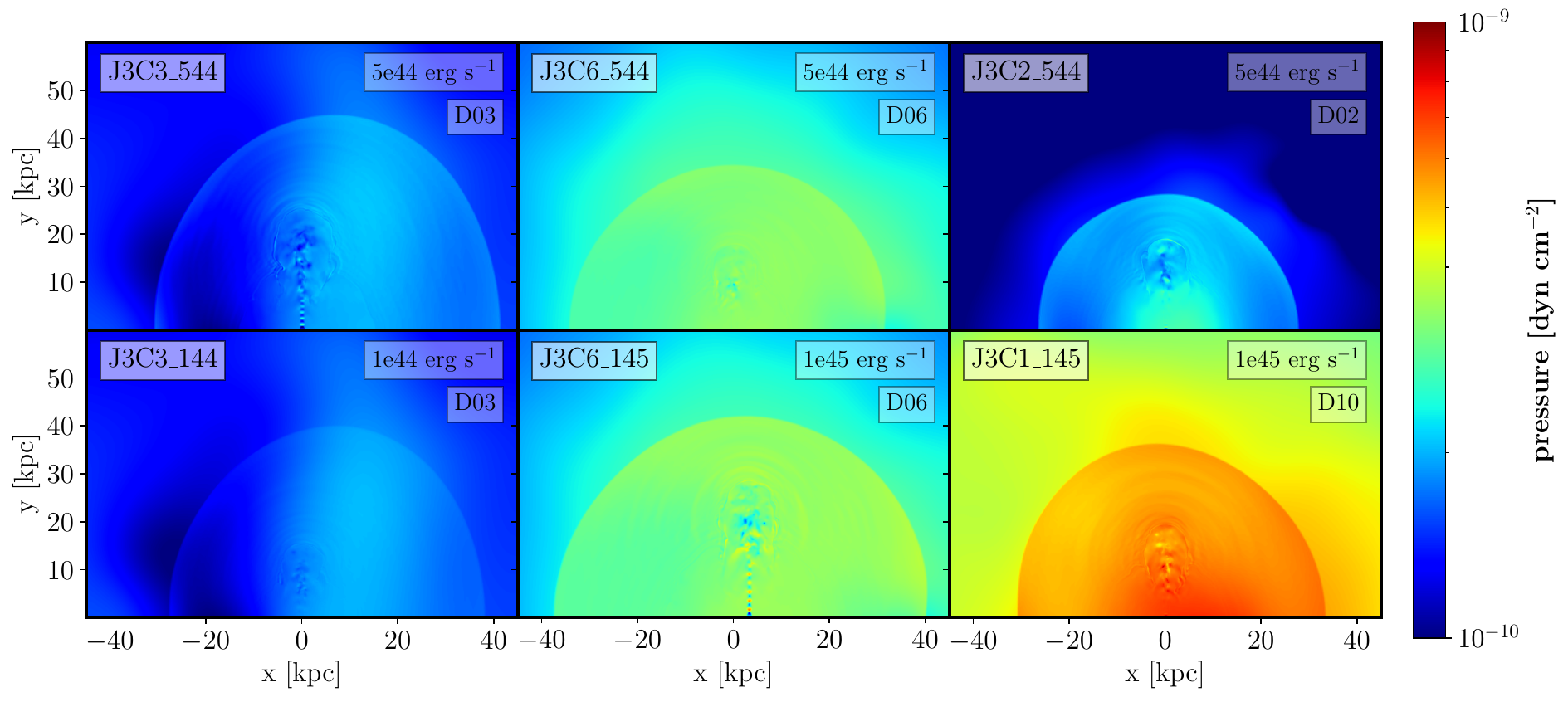}
\caption{Pressure maps of each simulation at the dynamical time t = 16 Myr, following the same order as in Fig. \ref{Fig_dens}.}
\label{Fig_pres}
\end{figure*}

\begin{figure}[htbp]
    \includegraphics[width=0.49\textwidth]{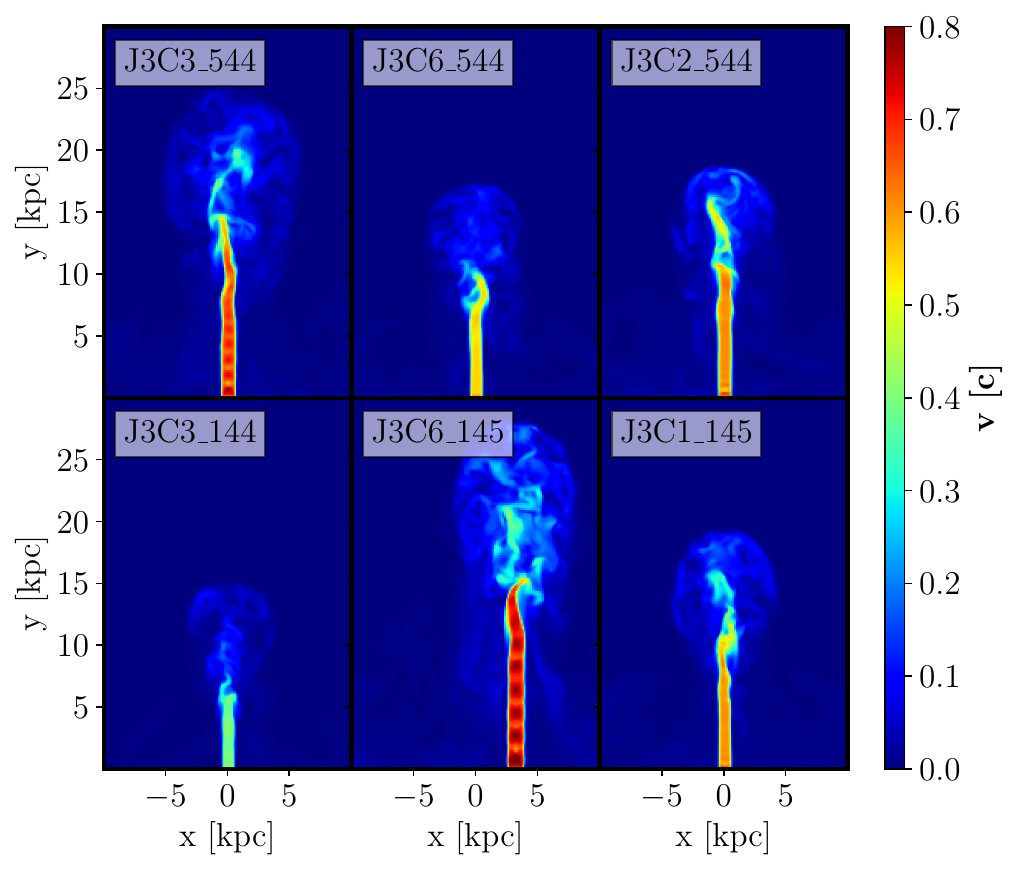}
\caption{Velocity maps of each simulation at the dynamical time t = 16 Myr, displayed in the same order as in Figs. \ref{Fig_dens} and \ref{Fig_pres}.}
\label{Fig_vel}
\end{figure}

\begin{figure}[htbp]
    \centering
    
   \begin{subfigure}[b]{0.45\textwidth}
       \includegraphics[width=\textwidth]{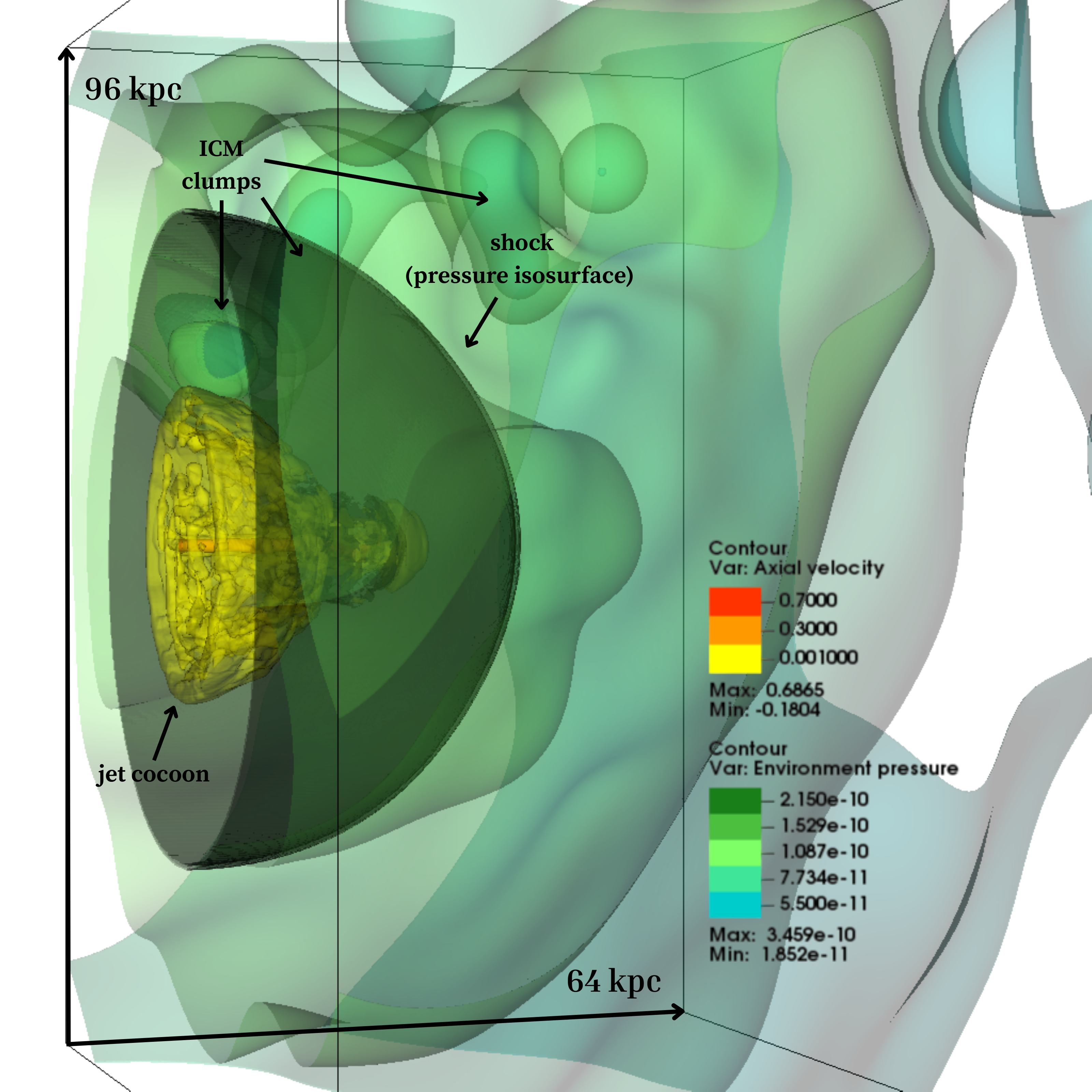}
   \end{subfigure}

   \vspace{1em}
   
   \begin{subfigure}[b]{0.45\textwidth}
       \includegraphics[width=\textwidth]{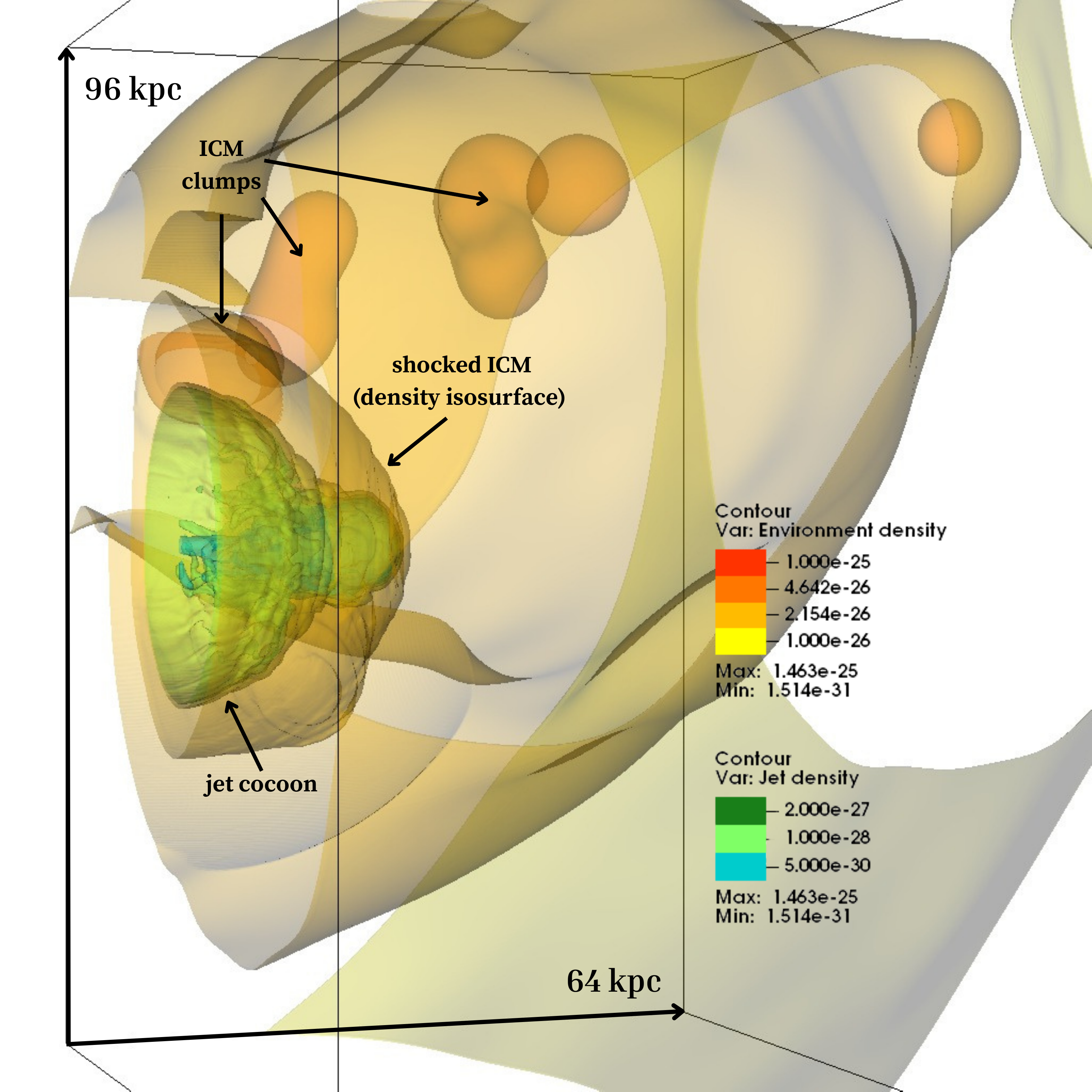}
   \end{subfigure}

\caption{3D model of jet simulation J3C2$\_$544 at 18Myr. \textit{Top}: Axial velocity isosurfaces (yellow-red scale) and ICM pressure isosurfaces (blue-green scale), included to show the jet cocoon and the shocked ICM. \textit{Bottom}: Density isosurfaces for the environment (yellow-red scale) and the jet (blue-green scale).}
\label{3D}
\end{figure}

\subsection{Large-scale structure}\label{sec_results_structure}

Figures~\ref{Fig_dens}, \ref{Fig_pres}, and \ref{Fig_vel} show 2D cuts of density, pressure, and velocity distributions, respectively, for all simulations at $t = 16\,{\rm Myr}$. Figure~\ref{Fig_dens} displays the density maps, where we can identify the different structures of the evolved jets (indicated in the central bottom panel for J3C6$\_$145).
The injected underdense material forms the central and collimated jet spine, which can be seen in the darkest blue, ending at the jet terminal shock. The jet spine plasma gets deflected at this reverse shock and forms the jet cocoon, which can be seen as the dark blue to green-yellow region surrounding the jet. Finally, the denser shell, in dark red, corresponds to the shocked ICM. The green-orange region tracks the mixing at the interface between the cocoon and the shocked ICM material.

When comparing the cases of jets with the same power that are injected in different environments (e.g., J3C3$\_$544, J3C6$\_$544, and J3C2$\_$544, all with $L_{\rm j}=5\times10^{44}~\rm{erg/s}$ jets in clusters D03, D06, and D02, in the upper row of Fig.~\ref{Fig_dens}), we see how denser environments result in slower jet propagation and consequent smaller and denser shocked shells. If we compare the jets with different injection powers evolving in a given cluster (e.g., J3C6$\_$544 and J3C6$\_$145, $5\times 10^{44}$ and $10^{45}\,{\rm erg/s}$ jets in galaxy cluster D06; or J3C3$\_$144 and J3C3$\_$544, $10^{44}$ and $5 \times 10^{44}\,{\rm erg/s}$ jets in galaxy cluster D03), we see how higher power jets expand faster, as expected. In some environments, as in cluster D03 or D02, the bow-shocks develop visible asymmetries, as a response to the substructures met by the expanding shock. The shock expands faster through the more dilute regions, and hence it generates a deeper rarefaction in its wake, which can eventually generate a global tilt of the jet/lobe in that direction. 

Figure~\ref{Fig_pres} shows 2D cuts of the pressure distribution for all simulations. We can observe the sphericity of the bow shock and the cocoon overpressure. The high sound speed in the post-shock region favors rapid homogenization of pressure, as also observed in higher power jet simulations \citep[e.g.,][]{perucho_long-term_2019,perucho_long-term_2021}. Still, the pressure peaks at the terminal shock region, where the jet kinetic energy is efficiently dissipated. Nevertheless, the environmental asymmetries and substructures are imprinted in the shocked ambient medium, which can develop significant inhomogeneities. Precisely because these depend on the ambient medium distribution prior to the passage of the shock, we observe similar features and pressures for the different jets that evolve through the same clusters (see, e.g., the left and central columns of Fig.~\ref{Fig_pres} for clusters D03 and D06, respectively). In all cases, the shocked region approaches pressure equilibrium with the ambient medium as jets expand, becoming barely distinguishable from the ambient medium in the color scale (mainly for clusters D03 and D06). 

Figure~\ref{Fig_vel} shows cuts of the $3-$velocity modulus ($\sqrt{v_x^2+v_y^2+v_z^2}$) distribution.
The panels show that the jet keeps mildly relativistic velocities down to the terminal shock, similar to the injection velocities chosen for our different models (see Table~\ref{tab_simulations}), which implies little or no deceleration from the injection to this point. Furthermore, we can also see how models J3C1$\_$145 and J3C2$\_$544 ($10^{45}$ erg s$^{-1}$ in D10 and 5$\times 10^{44}$ erg s$^{-1}$ in D02) give similar velocity distributions and sizes, even though the former has an injection speed of $0.6\,c$ and the latter of $0.4\,c$. This shows that denser environments, D10 in this case, can affect not only the development of the bow-shocks but also the jets, by indirectly contributing to the dissipation of kinetic energy via the generation of more frequent recollimation shocks or faster instability growth-rates.

The top panel of Fig.~\ref{3D} shows isocontours of axial velocity and pressure for simulation J3C2$\_$544. The velocity isosurfaces delineate the jet beam (in orange) and the cocoon (in yellow). In the cocoon, where two positive velocity contours are shown, the flow exhibits irregular features, as expected for a region filled with shocked, turbulent gas, including both backflowing (with a minimum axial velocity of $-0.18\,c$) and outflowing regions. The bow shock triggered by the evolution of the jet is traced in the image by the dark isosurface surrounding it. As anticipated by the 2D cuts, we see how the shock meets the clumps of the inhomogeneous ambient medium, traced by the pressure isosurfaces (in the blue-green color scale). The bottom panel shows isosurfaces of rest-mass density. One isosurface is chosen to show the shocked ICM region and shows a scaled-down version of the shock shape at larger scales, indicating that inhomogeneities in the environment can generate asymmetries in the distribution of shocked gas over large scales. The concentrations of gas in the ICM (clumps) can be easily identified with those in the top panel.

\subsection{Environment and shock coupling}\label{sec_results_anysotropy}

\begin{figure*}
    \centering
    \includegraphics[width=\linewidth]{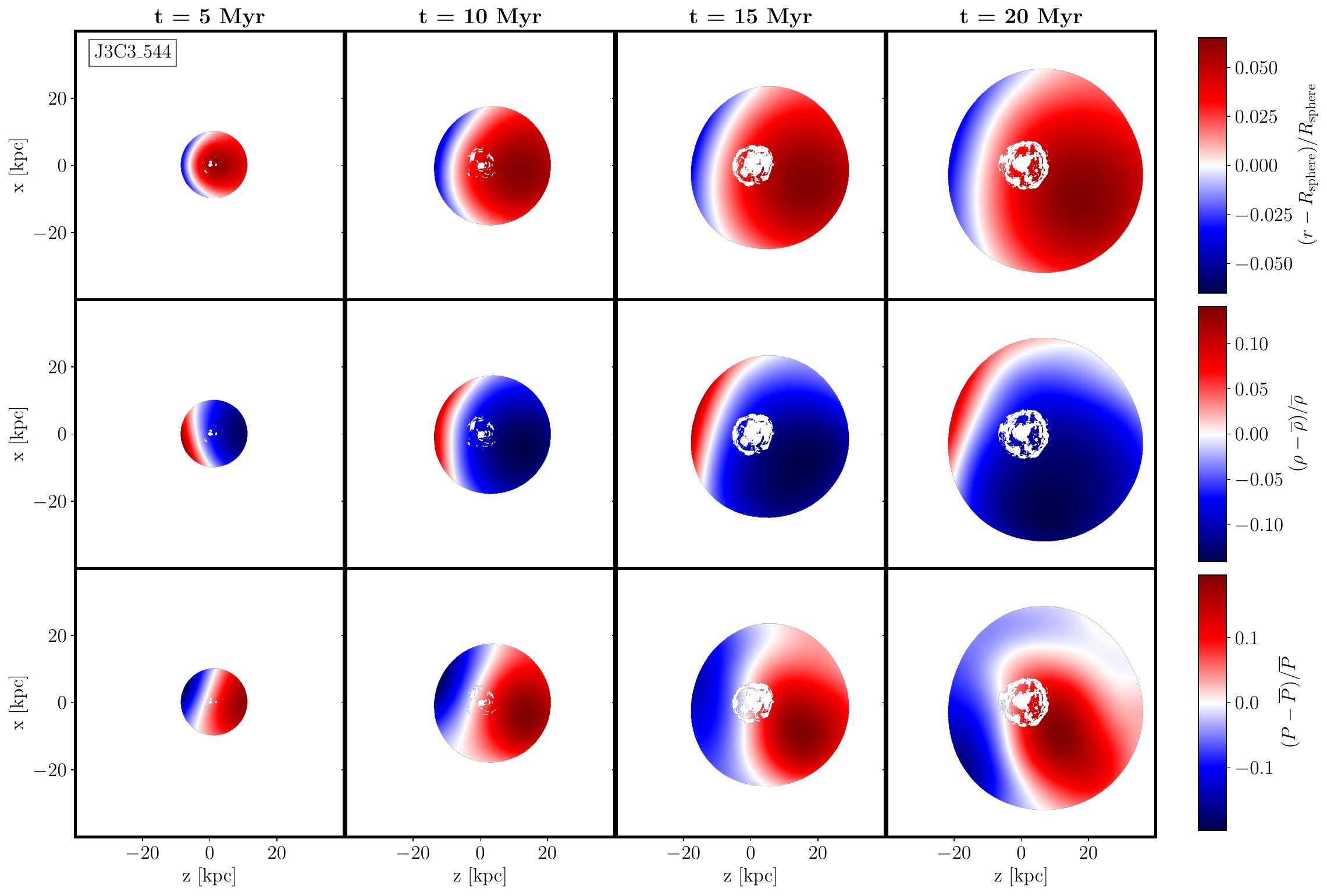}
    \caption{Time evolution of the deviation maps (with columns corresponding, from left to right, to 5, 10, 15, and 20 Myr) for J3C3$\_$544 simulations, illustrating the coupling between environment and jet propagation. Red (blue) regions denote positive (negative) deviations. \textit{Top row}: Relative deviations of the bow shock surface from an idealized spherical expansion. \textit{Middle and bottom row}: Relative ICM density and pressure fluctuations, respectively, at the bow shock surface.}
    \label{fig:asymmetries}
\end{figure*}

Our simulated jets propagate through the ICM clumps, leading to bow shocks with significant structural asymmetries, as illustrated by Fig.~\ref{3D}. To quantify the degree to which such inhomogeneities shape the shock propagation, we evaluated the deviations from an idealized spherical expansion of the shock front, alongside the local fluctuations of the unshocked ICM density and pressure. We present the resulting deviation maps in Fig. \ref{fig:asymmetries}, for the particular case of simulation J3C3$\_$54 at different evolutionary times (see the correlation maps for the remaining simulations at $t = 15\text{ Myr}$ in Appendix~\ref{appendixB}).

To derive these maps we first computed the shock front radial distance from the injection point, for the 3D evolved simulation data at each chosen time, by identifying the shock position for each transverse column across the grid:
\begin{equation}
 r(x, z) = \sqrt{(x - x_0)^2 + y_{\rm shock}^2(x, z) + (z - z_0)^2}
.\end{equation}
The reference idealized shock radius is then defined as the mean radial distance averaged over the detected shock surface: $R_{\rm sphere} \equiv \langle r(x, z) \rangle_{\rm shock} $. Hence, the local relative radial deviations of the shock front, presented in the first row maps of \ref{fig:asymmetries}, are computed across the shock surface:
\begin{equation}
    \frac{\delta r(x,z)}{R_{\rm sphere}}= \frac{r(x, z) - R_{\rm sphere}}{R_{\rm sphere}}
.\end{equation}

Secondly, to measure the environment density and pressure encountered by the shock, the unperturbed density $\rho_0(x, y_{\rm shock}, z)$ and pressure $P_0(x, y_{\rm shock}, z)$ are sampled from the initial cluster configuration (at $t = 0$) at the coordinates of the detected shock front. The reference idealized values for these parameters are also defined as the mean averaged across the shock surface:
\begin{equation}
        \bar{\rho} \equiv \langle \rho_0(x, y_{\rm shock}, z) \rangle_{\rm shock}, \qquad \bar{P} \equiv \langle P_0(x, y_{\rm shock}, z) \rangle_{\rm shock}.
    \end{equation}
The local relative ambient fluctuation maps, shown in the first row of Fig. \ref{fig:asymmetries}, are then derived from    \begin{equation}
        \frac{\delta \rho(x, z)}{\bar{\rho}} \equiv \frac{\rho_0(x, y_{\rm shock}, z) - \bar{\rho}}{\bar{\rho}}, \qquad \frac{\delta P(x, z)}{\bar{P}} \equiv \frac{P_0(x, y_{\rm shock}, z) - \bar{P}}{\bar{P}}.
    \end{equation}

To statistically evaluate the relationship between the cluster substructures and the shock morphology (while removing local cell-to-cell numerical autocorrelation), the transverse $(x, z)$ plane is partitioned into 16 independent azimuthal sectors, where we calculated the azimuthally averaged profiles $\langle \delta r / R_{\rm sphere} \rangle_\phi$, $\langle \delta \rho / \bar{\rho} \rangle_\phi$, and $\langle \delta P / \bar{P} \rangle_\phi$, and evaluated the Spearman rank correlation coefficient (a nonparametric measurement of the strength and direction of the statistical dependence between two variables), $\rho_{\rm sp}$, for two distinct pairs of deviations: the shock radius versus the ambient density, and the shock radius versus the ambient pressure.

We find a persistent spatial anticorrelation between the bow shock surface deviations and the local ICM density fluctuations in all simulations, with $\bar\rho_{\rm sp} (\langle \delta r / R_{\rm sphere} \rangle_\phi$, $\langle \delta \rho / \bar{\rho} \rangle_\phi)$ ranging from $- 0.39$ to $-0.98$ (and a mean value of $-0.84$) and reaching high significance ($p_{\rm val} < 3\times 10^{-6}$) in five of the six runs. As the shock propagates farther into low-density regions it can expand faster, while it is slowed by high-density clumps. This behavior establishes that the bow shock morphology acts as a record of the ICM density structure through which the jet has expanded.

Regarding the second pair (the bow shock radius and the ICM pressure), an apparent correlation is also found, with comparable strength values for four out of the six simulations ($\bar\rho_{\rm sp}(\langle \delta r / R_{\rm sphere} \rangle_\phi,\langle \delta P / \bar{P} \rangle_\phi) = (-0.24,-0.82)$, with $p_{\rm val} < 0.36$). However, the direction of the pressure anticorrelation is not consistent across the whole set: simulations J3C3$\_$544, J3C3$\_$144 show instead positive correlations ($\bar\rho_{\rm sp} = (0.96-0.99)$, with $p_{\rm val} < 3\times 10^{-9}$). This suggests that, while the ICM pressure structure could leave imprints on the bow-shock shape in individual environments, its effect cannot be generalized in the same way as for density distributions.

\subsection{Heating of the intracluster medium}\label{sec_results_heating}

\begin{figure}[htbp]
    \centering
    \includegraphics[width=0.49\textwidth]{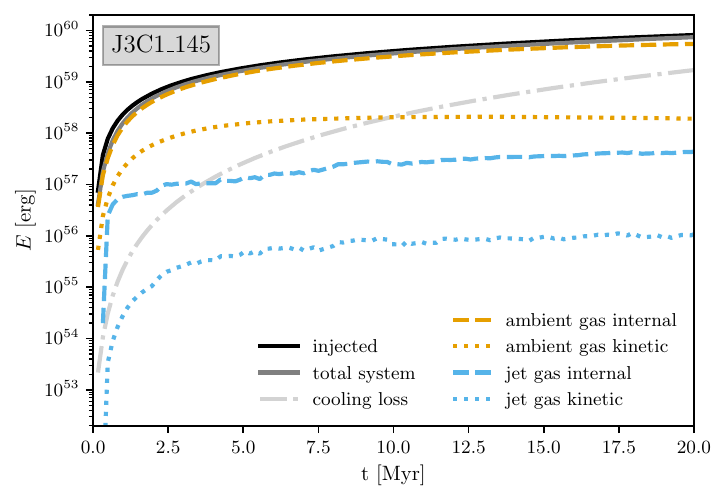}
\caption{Time evolution of the energy for the simulation J3C1 145, with a jet power of $10^{45}\,{\rm erg/s}$ in the D10 galaxy cluster. The solid black line represents the total energy injected by the jet; it coincides with the solid gray line, which corresponds to the total energy acquired by the system. The dashed and dotted orange lines correspond to the internal and kinetic energy of the ambient particles, respectively; blue lines correspond to same energies but for the particles injected with the jet.}
\label{energetics}
\end{figure}

Figure~\ref{energetics} shows the time evolution of the total injected energy (black line) and the internal (dashed lines) and kinetic (dotted lines) energies gained by the ambient medium (in orange) and kept by the jet flow component (in blue) for simulation J3C1$\_$145. We also show the increase in the total energy in the shocked region (dark gray line), obtained by adding the internal and kinetic energies gained by the medium and stored in the jet components, and the losses due to cooling (light gray dash-dotted line). The equivalent evolution plots for the rest of simulations are shown in Appendix~\ref{appendixA}.

The corresponding energy increments are computed according to the relativistic expressions. The total energy in a cell can be written as
\begin{equation}
    E_{\rm tot}\, =\, \rho_i(t) \Gamma_i(t) \left[ h_i(t) \,\Gamma_i(t) - p_i(t) - 1)\right] c^2 V_i,
\end{equation}
where $\rho_i(t)$, $p_i(t),$ and $h_i(t)$ are the gas rest-mass density, pressure, and specific enthalpy ($h=1+\varepsilon+p/\rho$, with $\varepsilon$ the specific internal energy), respectively, and $\Gamma_i$ the Lorentz factor \citep[see][]{perucho_long-term_2019} at a certain cell $i$ of volume $V_i$, and at a given time, $t$. Making use of the ideal gas equation of state with a variable adiabatic index (the Synge equation of state), which implies that pressure is given by $p=(\gamma - 1)\rho \varepsilon$, with $\gamma$ the adiabatic index, we derived the following expressions for the internal and kinetic energies in the ambient medium (subtracting the original value at the cell) and the jet:

\begin{eqnarray}
    \nonumber
    \Delta E_{\rm int,a}(t) & = & \sum_{i=1}^{N_{\rm cells}} \rho_i(t) \varepsilon_i(t) c^2 \Big[ \gamma_i(t) \Gamma_i(t)^2 - (\gamma_i(t)-1)\Big] \big(1-f_i(t)\big) V_i \\
    & & - \sum_{i=1}^{N_{\rm cells}} \rho_{i,0} \varepsilon_{i,0} \,c^2 V_i\,,
\end{eqnarray}
\begin{equation}
    \Delta E_{\rm kin,a}(t) = \sum_{i=1}^{N_{\rm cells}} \rho_i(t)\Gamma_i(t) c^2 \Big(\Gamma_i(t)-1\Big) \big(1-f_i(t)\big) V_i\,,
\end{equation}
\begin{equation}
    \Delta E_{\rm int,j}(t) = \sum_{i=1}^{N_{\rm cells}} \rho_i(t) \varepsilon_i(t) c^2
    \Big(\gamma_i(t) \Gamma_i(t)^2 - (\gamma_i(t)-1)\Big) f_i(t) V_i\,, 
\end{equation}
\begin{equation}
    \Delta E_{\rm kin,j}(t) = \sum_{i=1}^{N_{\rm cells}} \rho_i(t)\Gamma_i(t)c^2\Big(\Gamma_i(t)-1\Big) f_i(t) V_i.
\end{equation}

\noindent
In these expressions, $f_i(t)$ is the jet tracer in the cell at time $t$. The summation extends over the total number of computational cells, $N_{\rm cells}$. Note that since the environment is in approximate hydrostatic equilibrium, the increment in kinetic energy of the ambient gas is simply its kinetic energy. Finally, let us mention that the resulting energies are close to those obtained with the corresponding classical equations, since the relativistic effects are only important in the central jet spine, which involves a very small volume.

The plot shows that the total energy obtained for the shocked region is consistent with the energy injected by the jet. This helps confirm our results in terms of the energy conservation in the system. The largest portion of the injected energy is invested into the internal energy gained by the ambient medium, followed by the increase in kinetic energy as it is pushed outward. The former represents $\sim$80\% of the injected energy throughout the whole simulation (for all the simulations, see Appendix~\ref{appendixA}). This is one of the main results of this paper, as we show that even weak shocks are very efficient in transferring energy to ambient medium, as was already claimed for the case of stronger shocks driven by powerful, relativistic jets from numerical simulations \citep{perucho_intracluster_2011,perucho_large-scale_2014,perucho_long-term_2019} and analytical work \citep{perucho_radio_2017}. Actually, we can see that the efficiency is slightly reduced toward the end of the simulations because cooling losses become more and more relevant up to representing a 10\% of the injected energy by the end of the simulation (this explains the drop in the shocked ambient temperature; see Fig.~\ref{shocked_ICM}). 

\section{Discussion}\label{sec_discussion}

\begin{figure*}[htbp]
    \centering
    \includegraphics[width=0.8\textwidth]{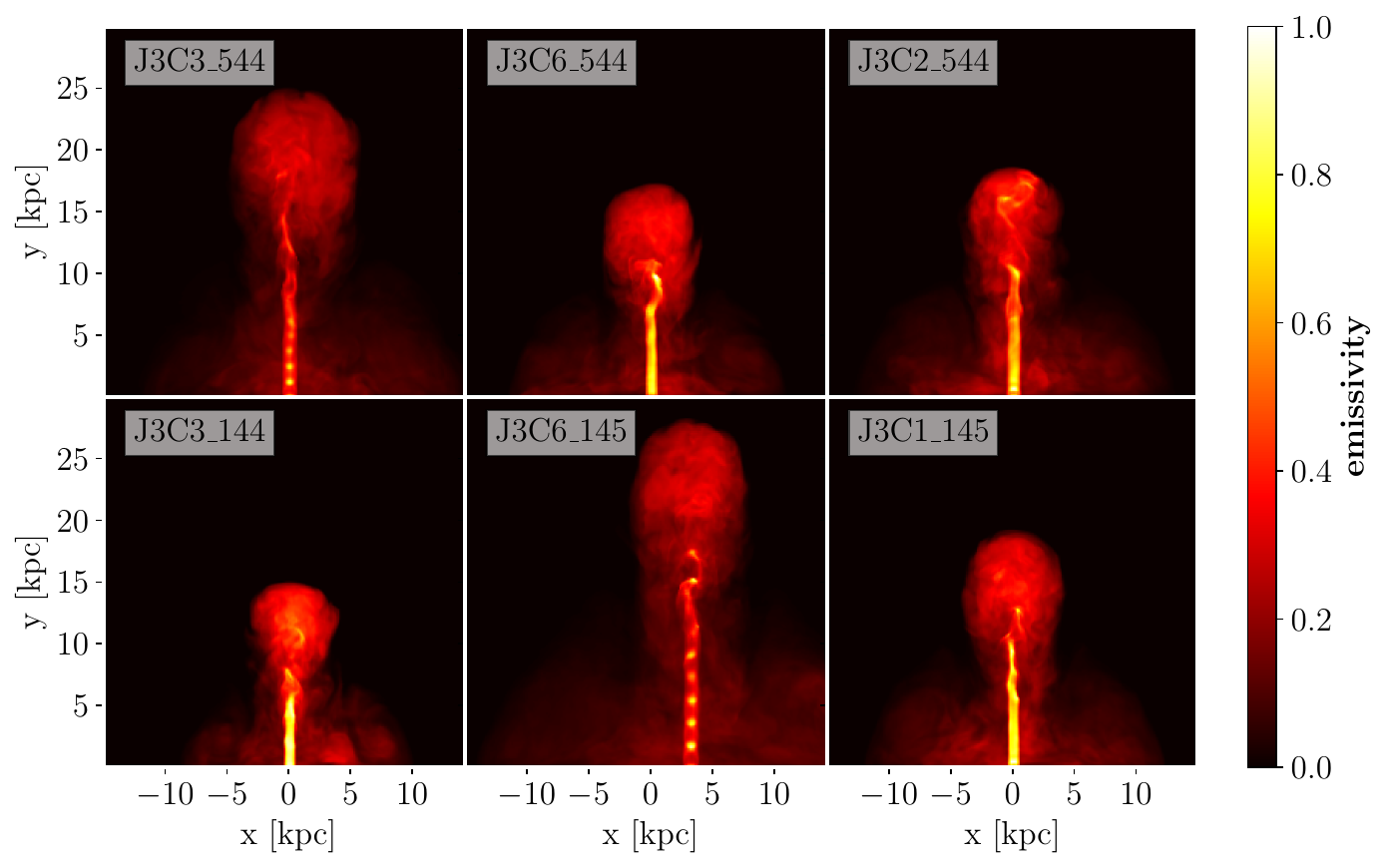}
\caption{Pseudo-synchrotron emissivity maps (in arbitrary units) computed according to Eq. 3.1 from \citet{perucho_numerical_2023}, for the multiple jet simulations at the dynamical time t = 16 Myr.}
\label{emissivity}
\end{figure*}

\subsection{Jet dynamics and propagation}\label{sec_discussion_dynamics}

The galaxy cluster profiles (see Fig.~\ref{Dianoga_profiles}) that we used as setups show pressures around one order of magnitude larger than those in the ambient media used in previous works for powerful jets 
\citep[e.g.,][]{perucho_large-scale_2014,perucho_long-term_2019}. The limited resolution achievable in 3D simulations requires the adoption of an initial jet radius that, while compatible with the observed opening angles of FRI jets \citep{laing_systematic_2014}, constrains the jet pressure at injection for a given jet power. As a consequence, the initial jet overpressure is reduced with respect to previous works, with an average value of 1.2 (except for simulation J3C3\_544, for which it reaches 3.6).

Once injected into the grid, the jet flow becomes collimated by the over-pressured cocoon, and all models develop the corresponding series of recollimation shocks (see, e.g., the top-left and bottom-central panels of Figs.~\ref{Fig_pres} and \ref{Fig_vel}). However, the relatively large initial jet radius (accounting for expansion from the launching region) also leads to a slower jet propagation, as discussed in \citet{perucho_radio_2017}. The influence of the jet opening angle on the deceleration of the jet head has been reported in previous works \citep{perucho_large-scale_2014,english_numerical_2019,stewart_cosmodragon_2025}. As a result, the bow shocks are quasi-spherical (see Figs.~\ref{Fig_dens}, \ref{Fig_pres}, and \ref{Fig_vel}), except for local inhomogeneities \citep[see also][]{yates_observability_2018}.

In an analogous way, although the jets in our simulations are initially supersonic, they quickly decelerate \citep{bicknell_model_1984,laing_systematic_2014} and become weakly supersonic after $2-3$ Myr of evolution, as also reported by \citet{stewart_cosmodragon_2025}, and in contrast with the aforementioned high-power jet simulations \citep[e.g.,][]{perucho_large-scale_2014}. The bow shock Mach numbers derived from our simulations ($\sim2$) are similar to those from observational estimates \citep{mcnamara_heating_2005,simionescu_large-scale_2009,croston_large-scale_2011,stawarz_interaction_2014}. We find an even better agreement, $M\approx 1.2$, if we estimate the Mach number by comparing the internal cocoon pressure with the ICM pressure at the jet head \citep[as suggested by][]{worrall_multiwavelength_2006}.

Despite evolving toward pressure balance \citep[as found for FRI observations by][]{croston_particle_2018}, a pressure mismatch between the shocked and the unperturbed ICM persists along the whole simulated time for all our simulations. Hence, the cavity expansion is still driven by weak shocks \citep[which is also consistent with][]{stewart_cosmodragon_2025}. We should run longer simulations to reach the transition to subsonic expansion, where buoyant expansion could take over. This transition has not been observed in numerical simulations of relativistic jets so far, but it has been suggested that it could eventually take place beyond the end of the active phase \citep[e.g.,][]{perucho_intracluster_2011,stewart_cosmodragon_2025}.

Figure \ref{emissivity} shows pseudo-synchrotron emissivity maps \citep[computed using Eq.~(3.1) from ][]{perucho_numerical_2023} of our simulated jets as viewed at $90^\circ$. The jets develop lobes that resemble the observable plume-like characteristic FRI morphologies: a bright region near the injection that fades toward the lobes. However, as in many other previous simulations of FRI jets, the deceleration occurs mainly via the development of a strong terminal shock driven by instabilities \citep[e.g.,][]{perucho_numerical_2007,massaglia_making_2016}, in contrast to the continuous process suggested by the analysis of radio emission from FRI jets \citep{laing_systematic_2014}. This result supports the idea that the environmental properties can play a significant role in determining the large-scale jet morphology (and FR classification), mainly at intermediate jet powers \citep{gopal-krishna_fanaroff-riley_2001,croston_environments_2019,mingo_revisiting_2019,stewart_cosmodragon_2025}.

\subsection{The role of the environment}\label{sec_discussion_icm}

While high-resolution hydrodynamic simulations of jets have traditionally modeled the ICM as an idealized smooth environment with pressure and density radially decreasing profiles 
\citep[e.g.,][]{hardcastle_numerical_2013,massaglia_making_2016,yates_observability_2018,perucho_long-term_2021}, observations reveal a significantly more complex ICM \citep{fabian_hst_2016,russell_alma_2017}. Our work follows a shift toward more realistic setups of NR jet simulations, adopting initial conditions from modeled ICM, including substructures and turbulent media, either added \citep[e.g.,][]{ehlert_self-regulated_2022} or emergent \citep[e.g.,][]{sotira_impact_2025,sotira_cold_2026,jlassi_simulating_2026}. Some such setups even simulate jets within “live” cosmologically evolved galaxy cluster environments \citep{heinz_answer_2006,mendygral_mhd_2012,bourne_agn_2019, bourne_agn_2021}. We bridged the gap between these approaches and relativistic jet physics by applying a relativistic solver to these realistic and cosmologically derived environments \citep[as in][]{yates-jones_cosmodragon_2023}, characterized by inhomogeneities and high density clumps.

Kiloparsec-scale simulations show that the interaction between jets, or their induced shocks, and ISM inhomogeneities can impact their global morphology and dynamics, as well as how effectively the jet couples to the surrounding medium \citep{wagner_relativistic_2011,mukherjee_relativistic_2018,talbot_blandfordznajek_2022}. From numerical simulations of NR jets evolving on cluster scales, it has been reported that cold and dense gas can inhibit jet propagation \citep{bourne_agn_2019,yates-jones_cosmodragon_2023} or induce deflections \citep{mendygral_mhd_2012,ehlert_self-regulated_2022}. Furthermore, recent high-resolution simulations have highlighted that jet interactions with multiphase media (either already present in the cluster or emerging due to previous activity phases) can produce irregular and asymmetrical jet or lobe morphologies \citep{bourne_agn_2021,sotira_cold_2026,jlassi_simulating_2026}. In general, these simulations involve large timescales (tens of gigayears), where the distorting interaction comes from the lobe material and the ambient, once the bow-shock has disappeared. However, in our work, we observe that inhomogeneities can arise in a much earlier stage (tens of megayears) at the shock interaction with the inhomogeneities met in the ambient media, even if they do not leave observable imprints on the current red jet and lobe system.

The overdense clumps of gas similarly reshape the bow shock and could eventually contribute to the mentioned flow deviations in longer timescales (see, e.g., the top-left panels in Figs.~\ref{Fig_dens} and \ref{Fig_vel}, where hints of jet/lobe deviation are already observed). The ICM density variations within the inner 40 kpc of the cluster core—a distance corresponding to the average jet extent at the end of our simulations—exhibit mean amplitudes of $\delta\rho / \rho \approx 0.18 - 0.34$, defined relative to the radially averaged background. These ICM fluctuations are inherent to the cosmologically derived environments, and they are the primary drivers of the observed cocoon asymmetries: the jet propagates faster through the lower-density local regions while encountering higher densities at dense clumps, which naturally leads to the non-axisymmetric shock growth shown in Fig.~\ref{Fig_dens}. From the correlation between the bow-shock asymmetry maps and the ICM density, illustrated in Fig.~\ref{fig:asymmetries}, we can confirm that it is the ICM density the one that shapes the bow-shock and cocoon/lobe propagation via ram pressure \citep{Begelman1989,marti_morphology_1997,Scheck2002}, rather than thermal pressure gradients, which show weaker correlation.

Beyond the localized influence of these inhomogeneities, the de-collimated morphology shown by our simulated jets, even for intermediate-to-high powers, is consistent with the idea of FRI sources being found, on average, in higher-density ICM than FRII sources \citep{gendre_relation_2013}. Such environments were thought to enhance jet confinement and deceleration, naturally leading to FRI-like morphologies. This was in good agreement with early classical studies suggesting a higher probability of finding FRI-like sources in richer cluster environments, as opposed to FRII sources, which were thought to reside in poorer environments or galaxy groups \citep{prestage_cluster_1988}. However, a more recent statistical analysis with large homogeneous samples \citep{massaro_deciphering_2019} found that both FR types live in galaxy-rich large-scale environments with similar richness, with the probability of finding an FRI lying in a dense neighborhood being higher than that for FRII sources (70\% versus 55\%).

Finally, X-ray observations suggest that the large-scale ICM structure is inherently turbulent \citep[see, e.g.,][]{sanders_constraints_2011,simionescu_constraining_2019}, likely due to structure formation and/or prior jet interactions. 
Achieving a self-consistent capture of how this environmental complexity affects long-term jet propagation and feedback requires full cosmological simulations. However, such models often need sacrificing spatial resolution and neglecting the relativistic regime \citep{schaye_eagle_2015,dubois_horizon-agn_2016,springel_first_2018}. Conversely, while significant efforts have been made to accurately resolve jets \citep{weinberger_simulating_2017,bourne_agn_2021} and/or incorporate relativistic effects \citep{yates-jones_cosmodragon_2023,stewart_cosmodragon_2025} in realistic clusters, the simultaneous inclusion of turbulence, cluster weather, and magnetic fields is key for a deeper understanding of feedback—features we defer to future work.

\subsection{Feedback impact from weak shocks on the ambient medium}\label{sec_discussion_feedback}

One of the main aims of the present work was to investigate the efficiency of intermediate-power jets [$5 \times 10^{44}-10^{45}$] erg/s and the weak shocks they trigger as a heating mechanism, in comparison with powerful jets \citep{perucho_large-scale_2014,perucho_long-term_2019,perucho_long-term_2021}. Our results corroborate a very efficient transfer of the injected energy into the ambient medium internal energy, with a small drop with respect to \cite{perucho_long-term_2019} (80$\%$ versus 95$\%$). This drop is probably related to the relatively low velocity and the large jet radius at injection and, at the late stages of the simulation, to cooling (Fig.~\ref{energetics}), which is more relevant in the low-power simulations.

From an observational perspective, low Mach number shocks surrounding the radio lobes in different objects have been reported \citep{mcnamara_heating_2005,simionescu_large-scale_2009,croston_large-scale_2011,laing_deep_2011,stawarz_interaction_2014}. Numerical simulations of NR jets have traditionally concluded that shocks are not very strong \citep{bruggen_shock_2007,li_modeling_2014,yang_how_2016,bourne_agn_2021,ehlert_self-regulated_2022} and thus are not the main feedback mechanism \citep[even for powerful jets in the NR regime; see][]{perucho_large-scale_2014}. In some cases, the authors conclude that only the strong shocks generated within the jet cone can play an important role or that weak shocks alone are unable to completely offset cooling \citep{yang_how_2016}.

However, from our relativistic simulations of jets with powers in the range $10^{44}-10^{45}\,{\rm erg/s}$ evolving in relatively high-pressure environments, we estimate that an 80$\%$ of the injected energy is used to increase the ambient medium internal energy. This implies that during 20~Myr of activity, a total of $6\times10^{58}$ ($\simeq 3.7\times10^{67}\,{\rm keV}$) to $6\times10^{59}\,{\rm erg}$ ($\simeq 3.7\times10^{68}\,{\rm keV}$) are injected into the central region of the cluster. Considering a region of $3\times10^{4}~{\rm kpc}^3 \simeq 10^{69}{\rm cm}^3$ and taking into account that the interstellar number densities are on the order of $0.01 - 0.1\,{\rm cm}^{-3}$, our results imply an energy budget per particle of around $1-10$~keV, way above the energy per particle required to stop cooling flows, according to \citet{mcnamara_heating_2007}. Moreover, we have seen that weak shocks evolve to produce quasi-spherical shapes, leading to an isotropic distribution of the injected energy. Hence, our results not only support the efficiency of weak shocks in heating the ICM but also show that they can provide the required energy to stop cooling flows. 

Therefore, we can claim that weak shocks can potentially be the main heating mechanism in FRI-like and/or low-power radio galaxies, at least for as long as these shocks do not disappear. \citet{laing_deep_2011} presented evidence of the presence of weak shocks surrounding the lobes of different FRI jets using deep imaging. We could expect this weak-shock phase to be longer, and thus more relevant, in intermediate power jets ($10^{44}-10^{45}\,{\rm erg/s}$).
At this range of powers we can probably find a large population of radio galaxies \citep{hardcastle_radio-loud_2019}, putting the weak shock heating at the center of cooling flow regulation and, hence, galaxy growth and evolution. Furthermore, this conclusion can be extended to old, high-power jets, which can keep efficiently heating the environment, even after the galactic activity is over, in the remnant phase \citep{perucho_large-scale_2014,stewart_cosmodragon_2025}. For even lower-power jets, $< 10^{43}~{\rm erg/s}$, the supersonic phase is expected to be shorter (see, e.g., \citealt{beltran-palau_agn_2025} for predictions using semi-analytical models of radio-galaxy evolution), reducing the effect of weak shocks to a smaller volume, probably within the host galaxy.

\subsection{Caveats and comparison with other works}

First and foremost, our estimate of $80\%$ heating efficiency is consistent with previous RHD jet simulations with stronger shocks \citep{perucho_dissipative_2019}. However, this range of values is significantly higher than NR hydrodynamic simulations \citep[e.g.,][]{yang_how_2016, bourne_agn_2019}, which report values around $\sim 20-40\%$. This discrepancy arises from fundamental differences between the regimes and in jet injection parameters. As demonstrated by \cite{perucho_large-scale_2014} and \cite{perucho_radio_2017}, the efficiency of shock-driven heating is critically dependent on injection parameters: relativistic jets with narrow opening angles maintain a higher jet-head pressure and long-lived bow shocks, delaying the transition from 1D ballistic propagation to 2D expansion. Furthermore, by retaining higher hotspot pressures, these jets maximize the shock-swept volume and the subsequent thermal coupling to the ICM (see also \citealt{english_numerical_2016} and \citealt{ weinberger_simulating_2017} for other arguments regarding the relation between cocoon pressure and energy transfer). In contrast, NR simulations tend to assume wider injection angles ($> 15^\circ$) or lower velocities \citep[e.g.,][]{bambic_efficient_2019, yang_how_2016}, which inherently favor a broader expansion but lower shock pressures. Thus, the lower efficiencies reported in turbulent, ``live'' environments likely represent a baseline for momentum-driven or poorly collimated feedback, whereas our results demonstrate that accounting for relativistic shock dynamics and jet collimation is essential to capture the full energetic potential of AGN feedback during the active growth phase.

Our results focus on the high-efficiency coupling by the forward bow shock during the active growth phase. Once the shocks disappear, the perturbations propagate as sound waves, which have been shown to maintain the energy transfer, albeit less efficiently (see, e.g., \citealt{bruggen_shock_2007}, \citealt{yang_how_2016}, \citealt{weinberger_simulating_2017}, \citealt{bourne_recent_2023}, and references therein). Sound-wave heating acts as a complementary process in the jet heating of clusters, distributing feedback energy at large radii and during the post-active stages of jet duty cycles \citep{bambic_efficient_2019,stewart_cosmodragon_2025}, during their quiescent evolution \citep[e.g.,][]{fabian_sound_2017, bambic_efficient_2019}. We suggest that these two mechanisms operate on different timescales and spatial ranges: shock heating dominates the active phase, while sound waves likely prevail at later epochs, once the previous phase is exhausted. Furthermore, the prominence of shock-driven heating in our work compared to sound-wave-dominated models, such as those of \cite{bambic_efficient_2019}, is again likely linked to differences in jet injection parameters. As mentioned, their larger opening angles ($15^\circ-25^\circ$) lead to a more rapid weakening of the bow shock and an earlier transition to the transonic regime. By contrast, our RHD simulations maintain higher post-shock pressures and momentum fluxes at the hotspot, allowing the bow shock to persist for longer, even when weak. This delays the onset of the sound-wave-dominated regime and sustains efficient shock heating of the ICM beyond the active phase \citep[see the discussion in][]{perucho_large-scale_2014,perucho_long-term_2019}.

\section{Summary and conclusions}\label{sec_conclusions}

We have presented six 3D RHD simulations of intermediate-power jets ($L_{\rm j} \sim$ $10^{44}$–$10^{45}$$\,{\rm erg\,s^{-1}}$) evolving for 20~Myr through four different realistic galaxy cluster environments taken from cosmological \textsc{GADGET}-3 runs. The simulations were designed to investigate the evolution of jets, their resulting morphology, and the mechanisms by which these jets transfer energy to the inner ICM. This study extends previous numerical efforts focused on the evolution and feedback processes of powerful jets \citep{perucho_long-term_2019,perucho_long-term_2021} to lower-power jets, with the aim to study the differences in the feedback mechanisms. This regime is particularly relevant owing to the much larger number of low-power sources expected to exist in the Universe \citep[e.g.,][]{sabater_lotss_2019}.

The main results obtained from these simulations are:
\begin{itemize}
    \item The jets were injected with low velocities and large radii, which correspond to the opening angles typical of FRI-like jets. However, as our jets still evolve within young, high-pressure cocoons, they are collimated. As a consequence of this process, the jets develop a series of shocks that result in jet disruption and produce the types of lobes typically observed in young FRI sources. In contrast, we expect FRI jet deceleration to take place via the development of small-scale instabilities \citep[][]{matsumoto_linear_2017,gourgouliatos_reconfinement_2018} or interactions \citep{perucho_triggering_2020}. 
    \item The jets develop quasi-spherical weak shocks, thus resulting in isotropic heating.
    \item Shocks evolving in denser environments decelerate earlier, resulting in shorter, more compact shocked shells compared to sparser and less massive clusters.
    \item Weak-shock heating is the dominant energy transfer mechanism in these FRI-like jets. 80\% of the injected jet energy efficiently transforms into ambient internal energy even when shock Mach numbers reach values of $\sim 2$, in contrast to earlier predictions that limited shock heating to high Mach numbers.
    \item The bremsstrahlung cooling significantly affects the shocked ICM shells, leading to a significant temperature decrease as it expands; this was not observed in previous studies of powerful FRII jets.
    \item Energy budget calculations reveal that these FRI-like jets transfer approximately  $1-10\,{\rm keV}$ per particle over 20~Myr, exceeding the threshold required to prevent cooling flows and quench star formation at these scales.
\end{itemize}

Given that intermediate-power FRI-like jets are more numerous than their powerful FRII counterparts, weak-shock heating efficiency could play a crucial role in the regulation of cooling flows and hence galaxy growth and evolution. The simulations presented here rely on a number of simplifying assumptions. In particular, the ICM is modeled as hydrostatic and nonturbulent, whereas observed clusters exhibit bulk motions, mergers, and complex velocity fields. Magnetic fields and associated processes (e.g., thermal conduction) are also neglected. All these factors can modify both jet evolution and energy transfer predictions.

Future work should therefore include relativistic magnetohydrodynamic simulations of jet evolution to study the possible role of magnetic fields and related processes in feedback. Improvements in the characterization of the ambient medium, such as the inclusion of dynamical effects or a turbulent magnetic field, could also improve the comparison between our simulations and real galactic environments. In addition, a comparison of the physics included in different codes, namely kinematic and thermodynamical relativistic effects, magnetic fields, and cosmic rays, could also contribute to a better understanding of this crucial process in galaxy and cluster evolution. 

Finally, capturing the full long-term evolution and large-scale morphology of FRI jets remains computationally expensive due to the slower expansion and longer timescales. Furthermore, resolving lower-power jets across large domains with fixed Cartesian grids is also particularly demanding. Future efforts should therefore prioritize more extensive simulations by employing exponentially decreasing or adaptive grids, which would allow jet propagation to be followed over larger distances and further into the passive regime. This would make it possible to track the shock evolution through the transition to the subsonic regime and potentially evaluate the heating efficiency of sound waves as compared to weak shocks.

\begin{acknowledgements}

The authors thank the anonymous referee for the constructive suggestions and also Stefano Borgani for the cosmological simulation data and the useful discussions. The simulations have been carried out in the Mare Nostrum computer from the \emph{Red Española de Supercomputación} and further analysis has been performed at the \emph{Servei d’Informàtica de la Universitat de València}. 

This work has been supported by the Astrophysics and High-Energy Physics program supported by the Spanish Ministry of Science  and Generalitat Valenciana with funding from the European Union NextGenerationEU (\texttt{PRTR-C17.I1}) through grant \texttt{ASFAE/2022/005}, by the Spanish Ministry of Science, \texttt{MICIU/AEI/10.13039/501100011033/ERDF/UE}, trough grants \texttt{PID2022-136828NB-C43} and \texttt{PID2025-168247NB-C43}, and by the Generalitat Valenciana through grant \texttt{CIPROM/2022/49}. 
JMM, SP and VQ also acknowledge support by the Agencia Estatal de Investigación Española (AEI; grant \texttt{PID2022-138855NB-C33}) and by the Ministerio de Ciencia e Innovación (MCIN) within the Plan de Recuperación, Transformación y Resiliencia del Gobierno de España through the project \texttt{ASFAE/2022/001}, with funding from the European Union NextGenerationEU (\texttt{PRTR-C17.I1)}.

\end{acknowledgements}

\bibliographystyle{aa}
\bibliography{references}

\begin{appendix}

\onecolumn

\section{Energy transfer and feedback}\label{appendixA}
\begin{figure*}[h]
    \centering
    \begin{subfigure}[b]{0.45\textwidth}
        \includegraphics[width=\textwidth]{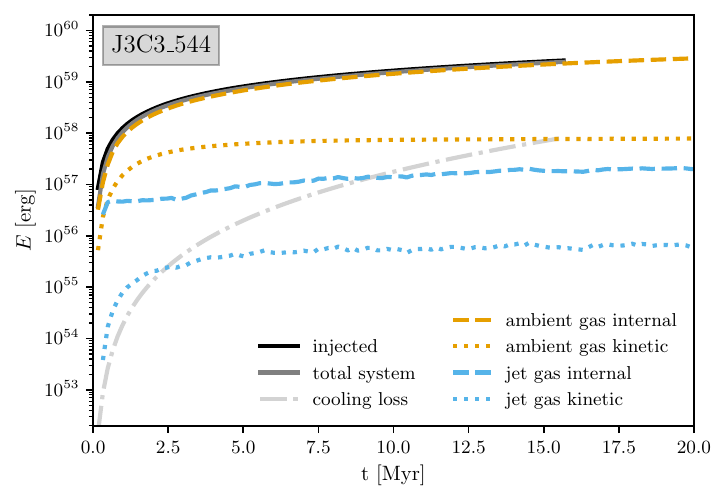}
    \end{subfigure}
    \hfill
    \begin{subfigure}[b]{0.45\textwidth}
        \includegraphics[width=\textwidth]{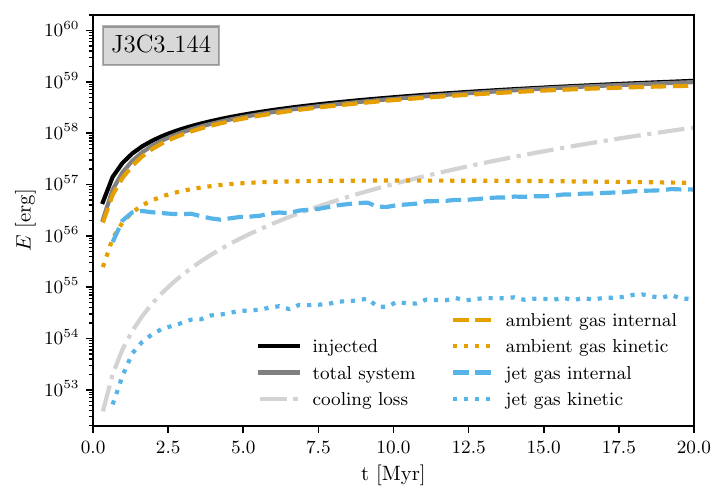}
    \end{subfigure}

    \vspace{1em}

    \begin{subfigure}[b]{0.45\textwidth}
        \includegraphics[width=\textwidth]{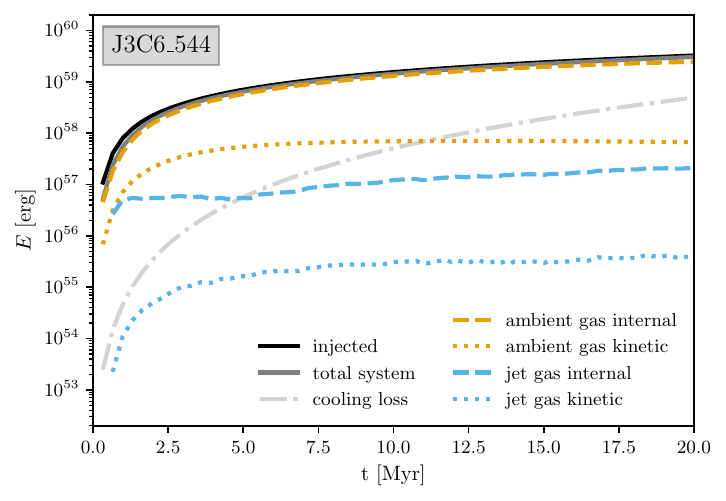}
    \end{subfigure}
    \hfill
    \begin{subfigure}[b]{0.45\textwidth}
        \includegraphics[width=\textwidth]{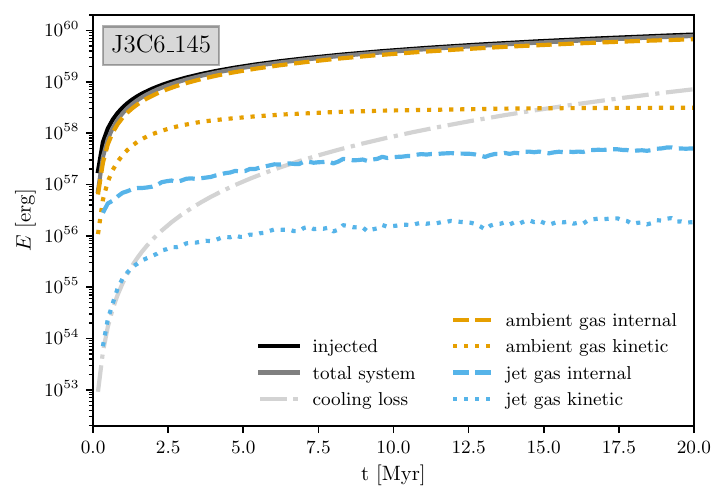}
    \end{subfigure}
    
    \begin{subfigure}[b]{0.45\textwidth}
        \includegraphics[width=\textwidth]{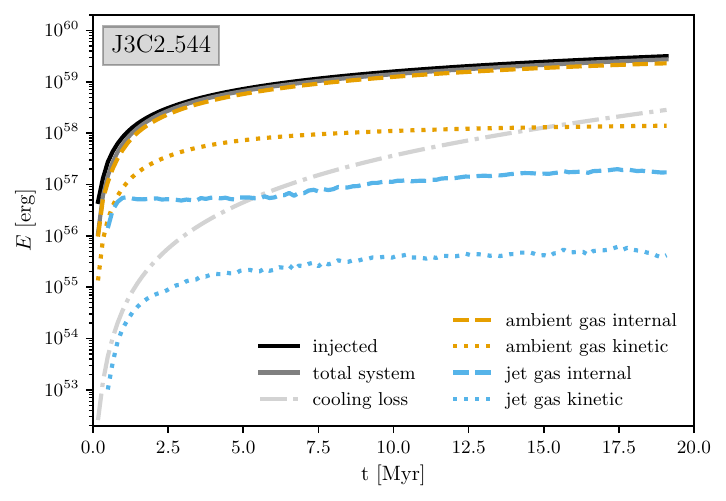}
    \end{subfigure}
    \hfill
    \begin{subfigure}[b]{0.45\textwidth}
        \includegraphics[width=\textwidth]{Figures/jet_energy_J3C1_145.pdf}
    \end{subfigure}

    \caption{Same as Fig. \ref{energetics} but for the remaining simulations. }
    
    \label{energetics_appendix}
\end{figure*}

\onecolumn

\section{Shock-environment coupling}\label{appendixB}
\begin{figure*}[h]
    \centering
    \begin{subfigure}[b]{0.78\textwidth}
        \includegraphics[width=\textwidth]{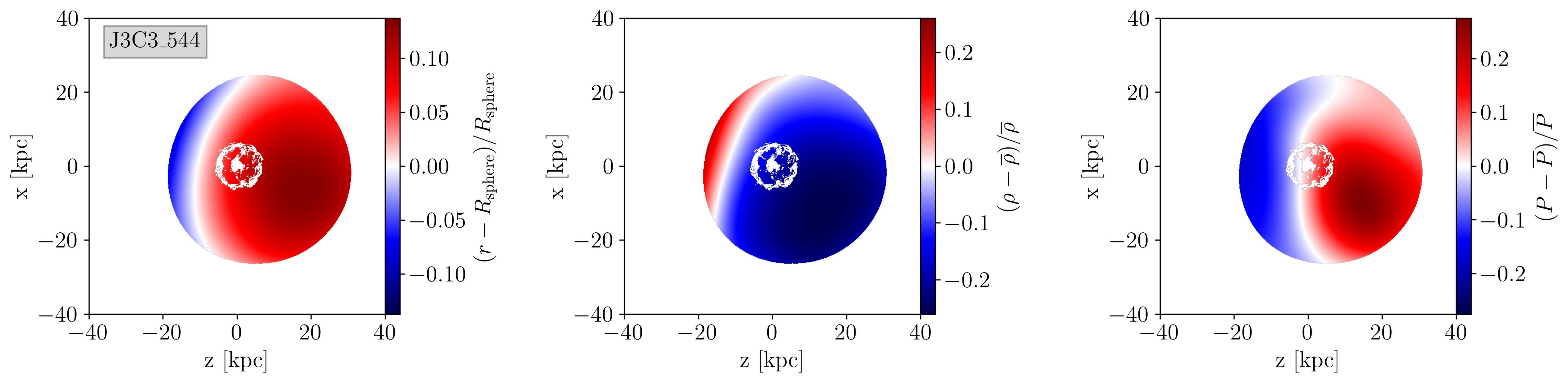}
    \end{subfigure}

    \begin{subfigure}[b]{0.78\textwidth}
        \includegraphics[width=\textwidth]{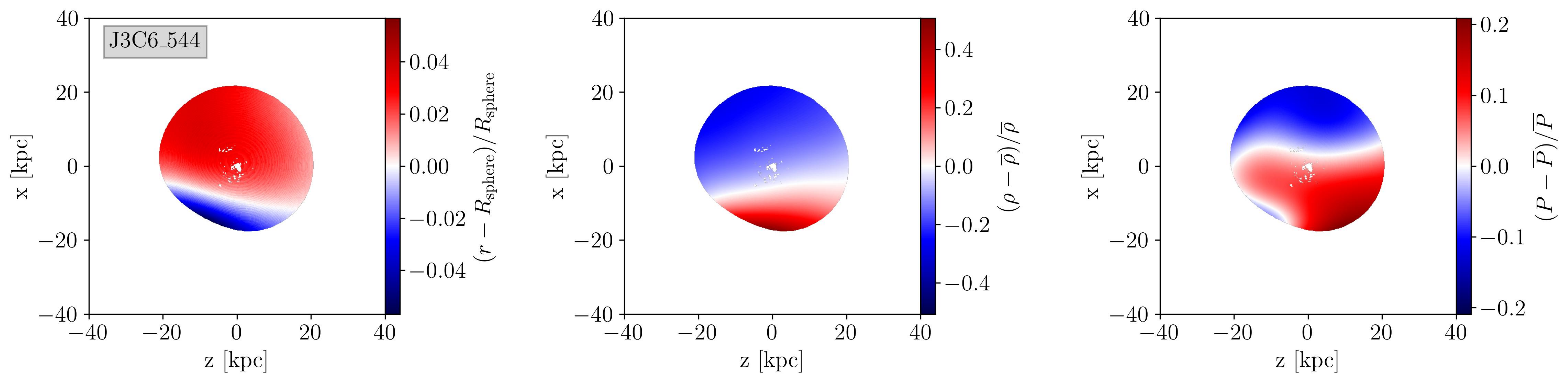}
    \end{subfigure}
    
    \begin{subfigure}[b]{0.78\textwidth}
        \includegraphics[width=\textwidth]{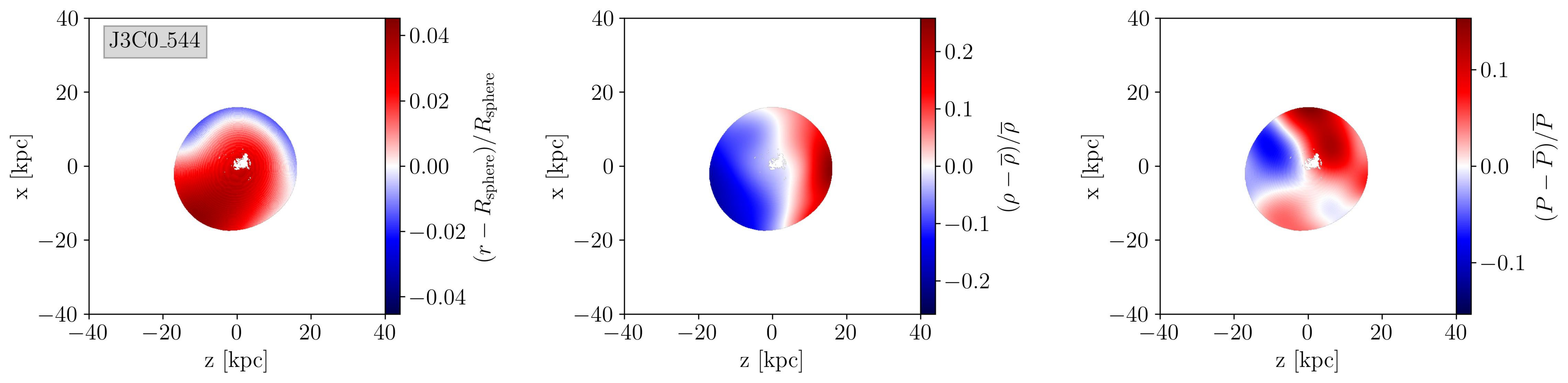}
    \end{subfigure}
    
    \begin{subfigure}[b]{0.78\textwidth}
        \includegraphics[width=\textwidth]{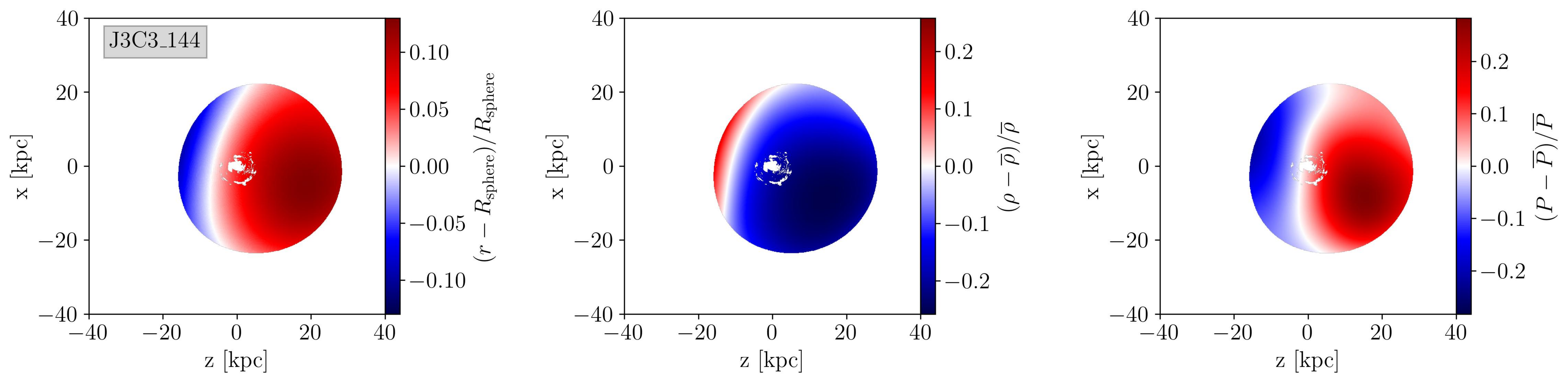}
    \end{subfigure}
    
    \begin{subfigure}[b]{0.78\textwidth}
        \includegraphics[width=\textwidth]{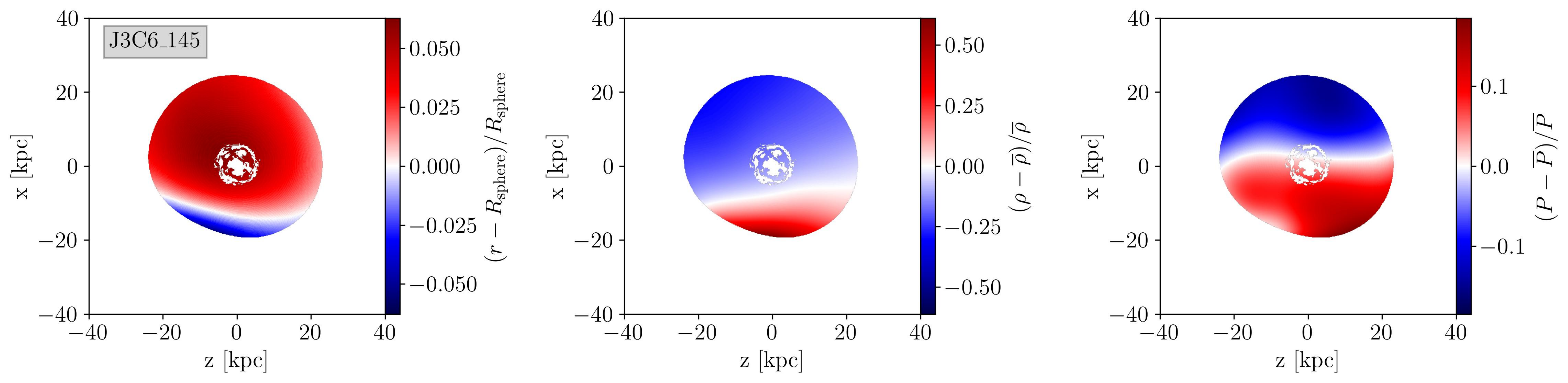}
    \end{subfigure}
    
    \begin{subfigure}[b]{0.78\textwidth}
        \includegraphics[width=\textwidth]{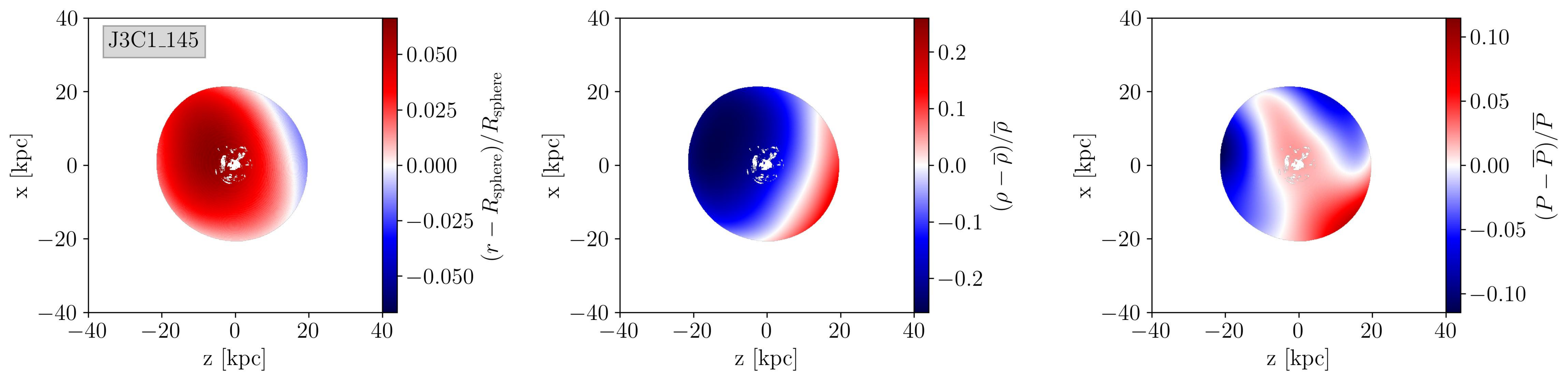}
    \end{subfigure}

    \caption{Deviation maps at 15Myr for all remaining simulations. \textit{Left to right}: Deviation maps of the bow shock surface and ICM density and pressure (at the bow shock positions), illustrating the coupling between the environment and bow shock radii. As in Fig. \ref{fig:asymmetries}, red (blue) regions denote positive (negative) deviations.}
    
    \label{asymmetries_appendix}
\end{figure*}

\end{appendix}

\end{document}